# Visual Matters: Connecting Aesthetic Appeal and Production Quality of Photos, Infographics and Data Visualizations to Credibility of Social Media Posts

Salman Khawar, University of California, Davis
Email: skhawar@ucdavis.edu
ORCID ID: https://orcid.org/0000-0002-7303-7774

Yingdan Lu, Northwestern University
Email: yingdan@northwestern.edu
ORCID ID: https://orcid.org/0000-0002-9955-6070

Yilang Peng, University of Georgia
Email: Yilang.Peng@uga.edu
ORCID ID: https://orcid.org/0000-0001-7711-9518

Jiyoung Yeon, University of Georgia
Email: jy.yeon@uga.edu
ORCID ID: https://orcid.org/0000-0002-6063-6452

Cuihua Shen, University of California, Davis
Email: cuishen@ucdavis.edu
ORCID ID: https://orcid.org/0000-0003-1645-8211

**Author Notes**

Declaration of conflicting interest: The authors declare no potential conflicts of interest with respect to the research, authorship, and/or publication of this article.

Funding statement: This study was supported by the National Science Foundation CNS-215076 and CNS-2150723.

Ethics considerations: This study was deemed as exempt from the Institutional Review Board of UC Davis (IRB ID: 1801813-3)

Correspondence concerning this article should be addressed to:
Cuihua Shen, University of California, One Shields Avenue, Davis, CA 95616, United States.
Email: cuishen@ucdavis.edu

**Abstract**

The rapid proliferation of visual content raises fundamental questions about how different visual formats and features shape perceived credibility. Drawing on processing fluency theory, this research examines how visuals shape credibility judgments. We focus on three popular formats—photos, infographics, and data visualizations—comparing them to text-only posts, and test how two visual features, aesthetic appeal and production quality, influence credibility through processing fluency as a mediating mechanism. Through a preregistered experiment with 1200 US participants, we found that visual posts are generally perceived as more credible than text-only posts but this credibility advantage only applies to photos and infographics, not to data visualizations. Aesthetic appeal increases perceived credibility, partially mediated by processing fluency, while production quality had no significant effect on credibility across formats. These findings differentiate visual formats, advance conceptualizations of visual features, and identify processing fluency as a key mechanism for theorizing credibility across multimodal contexts.

*Keywords: visual communication, aesthetic appeal, processing fluency, message credibility, social media*

**Visual Matters: Connecting Aesthetic Appeal and Production Quality of Photos, Infographics and Data Visualizations to Credibility of Social Media Posts**

Core communication constructs and processes such as credibility and persuasion were largely developed on textual and spoken stimuli, yet the contemporary online environment has become increasingly visual and multimodal. Every day, more than 3.2 billion images are uploaded to the internet (Thomson et al., 2020), and the share of US adults getting news regularly from Instagram increased from 28% to 41% between 2020 and 2025 (Pew Research Center, 2025). Yet visual credibility cannot be reduced to source or heuristic cues alone: the mere presence of visuals in online messages can enhance the credibility of information (Newman & Schwarz, 2024). Visual content, therefore, can have dangerous consequences, particularly through the spread of manipulated visuals and 'deepfakes' that are highly sophisticated synthetic media that convincingly replicate real people and objects. Citizens from certain demographic groups or those with limited digital media literacy are susceptible to such visual deception, which can have lasting effects on attitudes and beliefs (Weikmann & Lecheler, 2022). A core question for the field, then, is how the visual form of a message shapes audiences' credibility judgments independent of its source, platform, or content.

There are three gaps in the literature on visual credibility. First, most studies have considered non-visual factors such as source trustworthiness and virality cues (e.g., Shen et al., 2019), but have not focused on *visual* formats and features. The few studies that do examine visuals often treat them as undifferentiated stimuli, without considering how specific visual features are associated with human credibility perceptions. For example, Hameleers et al. (2020) compared multimodal versus textual messages containing disinformation, obscuring considerable variations among visuals themselves. In addition, most modality comparisons typically only focus on photos versus text, leaving non-photo visuals untheorized. Second, although visuals are often assumed to enhance credibility, we still lack a clear theoretical and experimental account of how specific visual features shape credibility judgments. Third, processing fluency theory offers a potential account of why visual and easy-to-process information feels true, but it has been built almost entirely on low-ecological-validity laboratory stimuli, such as font color and basic shapes (Reber & Schwarz, 1999). Whether and how fluency operates on ecologically valid, socially embedded visual messages remains an open theoretical question.

Addressing these gaps, this research develops and operationalizes the constructs of aesthetic appeal and production quality, and experimentally tests their effects on perceived credibility of visual social media posts. Extending existing visual research, which has focused primarily on photographs, we examined three visual formats—photos, infographics, data visualizations—through a preregistered online experiment ($N$=1200), and tested processing fluency as an underlying mechanism, hypothesizing that these visual features enhance perceived credibility of social media posts when the content is processed with less cognitive effort by audiences. Our findings highlight the general superiority of visual formats over text and the role of aesthetic appeal in enhancing the perceived credibility of social media posts. Theoretically, we provide a mechanism-level account of visual credibility that travels across communication subfields, including health science, political and journalism research, each of which increasingly relies on visual messages but lacks a shared theoretical understanding for them. Empirically, we extend fluency theory beyond impoverished laboratory stimuli and empirically identify boundary conditions on its predictive power. Practically, our findings can help social media platforms and its users better distinguish between credible and untrustworthy visual content by recognizing which visual features make posts more persuasive and potentially more likely to spread.

## Theory and Hypotheses

### Credibility of Visual Content

Credibility pertains to the audience's perception of the believability of information (Tseng & Fogg, 1999). Prior research has studied different kinds of credibility including medium credibility, pertaining to the specific medium used to convey information, and source credibility, pertaining to the perceived expertise or trustworthiness of the source (McDermott et al., 2024). This research focuses on message credibility, which is "an individual's judgement of the veracity of the content of communication" (Appelman & Sundar, 2015, p. 63). Specifically, we are interested in social media users' *perceived* credibility of visual posts, and not in the actual veracity of information presented in these posts.

Existing research on the credibility of visuals, especially in the disinformation context, can be divided into three strands. The first strand looks at differences in credibility perceptions between multimodal (photos combined with text) and text-only messages, finding the former to be generally more persuasive and more effective in spreading disinformation (Hameleers et al., 2020). The second strand approaches visual credibility from the perspective of social and heuristic cues. For example, Shen et al. (2019) tested the effect of whether visuals from trusted sources(e.g., news organizations) were rated as more credible than those from social media accounts and whether bandwagon cues, such as likes and shares, influenced credibility perceptions. The third strand looks at users' ability to detect deepfake videos (Jin et al., 2023) and manipulated images (Kasra et al., 2018). Findings from such studies suggest that online audiences often struggle to identify skillfully manipulated visuals and tend to rely heavily on non-visual cues such as source and popularity indicators, when making credibility assessments.

Prior research on visual credibility, however, does not look at how specific features of visual social media posts influence users' credibility perceptions. The current research aims to fill this gap by analyzing the influence of two important features of visual posts on credibility ratings. We use the theoretical framework of fluency theory to examine this relationship.

## Understanding Visual Credibility through Fluency Theory

Processing, or cognitive, fluency is the "experienced ease of ongoing mental processes" (Unkelbach & Greifeneder, 2013, p. 11). All cognitive activity, or mental processes, are accompanied by metacognitive feelings, which are subjective experiences that help in evaluating any such activity (Reber et al., 2004). Two main types of processing fluency are identified in prior research: conceptual fluency and perceptual fluency (Graf et al., 2017). Conceptual fluency is meaning-based and is experienced when a concept helps in processing another semantically related concept. For example, advertising studies have tested how presenting the target (product) in predictive contexts can enhance conceptual fluency and lead to more favorable brand evaluations (Lee & Labroo, 2003). Perceptual fluency, most relevant to the current research, is concerned with the perceptual features of a stimulus such as aesthetic features (Reber et al., 2004), visual clarity (Whittlesea et al., 1990), and complexity (Landwehr et al., 2011).

Fluency is a central metacognitive feeling that plays an important role in forming a range of human judgements. For example, statements that are easy to process are metacognitively experienced as being more familiar, and true, by individuals even if they have not seen or heard them before (Reber & Schwarz, 1999). Other judgments, in experimental research, pertaining to fluently processed stimuli include liking or preference for such stimuli, increased confidence when evaluating such stimuli, greater perceived intelligence of source, and fame of individuals depicted (Reber & Schwarz, 1999; Alter & Oppenheimer, 2009). The theory is based on the premise that less effortful cognitive tasks lead to more fluent, or positive metacognitive, experiences and subsequently positive judgements. Conversely, more effortful tasks lead to less fluent experiences and less positive judgments.

Prior research has explained the specific connection between higher processing fluency and increased credibility, also termed as the ‘truth value’ (Winkielman et al., 2012), using Brunswik’s lens model. According to the Brunswikian model (Brunswik, 1979), individuals use proximal cues, such as processing fluency, to make inferences about a distal criteria that are not directly accessible, such as truth or credibility. Through experience and feedback from their environment, humans learn that fluency and truth are positively correlated (Unkelbach & Greifeneder, 2013). Hence, when exposed to a new stimulus that is processed fluently, individuals are more likely to perceive it as true. This causal relationship between fluency and credibility has received empirical support. For example, Reber and Schwarz (1999) used different font colors of text against a white background to manipulate the visual ease with which participants could read a statement, and the results suggested that when the statement was more fluently processed it was perceived to be more true. It is therefore expected that users who experience processing fluency when viewing social media posts will evaluate their credibility favorably. In the following, we explicate three visual formats and two visual features tested in this research, from the theoretical lens of fluency theory.

**Visual Formats**

Visual formats refer to the different ways in which visual and multimodal data is presented. Some commonly used visual formats include photographic images, infographics, and data visualizations (Peng et al., 2023). The mere presence of visuals accompanied with textual messages has been observed to increase credibility of information; a phenomenon studied as the ‘truthiness’ effect (Newman & Schwarz, 2024). These judgments regarding the credibility of messages have been attributed to the ease of processing of information presented visually, which enhances the acceptance of the claims made in the accompanying text (Alter & Oppenheimer, 2009). Furthermore, visuals can even cause individuals to mistakenly draw inferences and reach false conclusions over and above information contained in text by inducing metacognitive feelings of familiarity, making the information feel truer (Smelter & Calvillo, 2020).

*Photographic images*, or photos, are static visual representations of reality. They are uniquely characterized by their true-to-life quality such that they directly depict physical objects and events as found in the non-mediated environment; a quality known as *indexicality*. In contrast, text is abstract as it constitutes symbols which require extracting semantic meaning and creating an imagined reconstruction requiring greater cognitive effort (Messaris & Abraham, 2001). Photos, even when being generic (but semantically related to text), help users understand and visualize elements of a claim contained in text (Newman & Schwarz, 2024). They may also serve as a memory cue making related information readily accessible helping in rapid intuitive judgements about credibility (Abed et al., 2017). Existing visual credibility research has predominantly focused on photos.

*Infographics*, or information graphics, are graphic visual representations of content, employing components such as charts, graphics and icons, used in fields such as journalism (Chun, 2023), health communication (Kong et al., 2024), and politics (Amit-Danhi & Shifman, 2020). They can serve multiple purposes such as conveying complex information and research findings, and encouraging behavioral change. The general preference of infographics over text can be attributed to the accessible and easy-to-process manner in which infographics convey complex information (Kong et al., 2024), but research is limited on the credibility of infographics. Li et al. (2018) found design quality of infographics to be positively associated with credibility perceptions of information presented. Other research either looks at credibility as a secondary outcome or considers credibility of the source rather than the message. For example, Kiernan et al. (2018) in an experiment on the effectiveness of infographics in promoting research literacy found that infographics led to greater trust in the source compared to text alone.

*Data visualizations* are a versatile and useful visual format to effectively convey complex statistical information in an engaging manner. They can range from simple graphs to complex data displays, and are used in multiple domains (Franconeri et al., 2021). While infographics generally curate and present results of data analyses to support information sharing and action, data visualizations include graphical representations of the data itself (Kong et al., 2024). Hence, data visualizations serve to uncover patterns in the data and present them intuitively, making processing of information easier, which would otherwise be difficult and time consuming to digest if relying on text alone. Moreover, the presence of graphs and visualizations increases persuasiveness of messages as they create a perception of being scientific (Tal & Wansink, 2014). Although limited in nature, research suggests that data visualizations could also enhance credibility perceptions. Henke et al. (2019) found that adding data visualizations to news stories increased participants' perceived message credibility, which was attributed to the ease of information processing and comprehensibility of data visualizations. Taken together, we predict that social media posts containing photos, infographics, and data visualizations will be perceived as more credible compared to text-only posts and also question whether the hypothesized credibility advantage applies consistently across visual formats.

***H1*: Social media posts with text and visuals (photos, infographics, data visualizations) will be perceived to be more credible than social media posts with text only.

***RQ1*: Is the positive effect of visuals over text on credibility consistently observed across visual formats (photos, infographics, data visualizations)?

## Visual Features

A visual feature pertains to a "special property of an image as a whole or an object within the image" (Weinmann, 2013, p.3). In the current research we consider two main visual features: aesthetic appeal and production quality.

### *Aesthetic Appeal*

Aesthetics refer to an artistically beautiful or pleasing appearance characterized by creativity and originality (Tractinsky et al., 2000). Aesthetic appeal is an abstract and complex concept studied from different perspectives (Lavie & Tractinsky, 2004). While classical studies of aesthetics have considered aesthetic properties to be inherent to objects, more recent studies involving information systems have adopted an interactionist perspective taking into account audiences' evaluations (Lazard & Mackert, 2015). To demonstrate the importance of this visual feature, a study conducted by Fogg et al. (2003) revealed that 46% of participants indicated the aesthetic appearance to be the most salient aspect when determining credibility of a website.

Beauty and truth are long considered two sides of the same coin (Reber et al., 2004). In other words, aesthetically pleasing stimuli are often perceived as more credible. This association can be explained by the idea that judgments of aesthetic appeal and credibility are shaped by the same underlying psychological mechanism, namely processing fluency. Experimental studies through manipulation of aesthetic features, including symmetry and contrast, of stimuli demonstrated that ease of processing enhanced credibility of messages (Reber & Schwarz, 1999; Reber et al., 1998). Therefore, aesthetic features facilitate the processing and identification of stimuli, which in turn leads to positive credibility evaluations (Winkielman et al., 2012).

Prior research in human computer interaction has established the positive influence of aesthetics of websites, and more recently social networking sites, on credibility judgements. Studies suggest that these judgments are made instantaneously, on a visceral level, in as little as 50 milliseconds (Robins & Holmes, 2008; Lindgaard et al., 2006). However, the influence of aesthetic appeal on credibility judgments has not been studied in the visual context, using visual social media posts specifically. We expect that aesthetically appealing visual posts will be perceived to be more credible than posts with low aesthetic appeal. We also test whether aesthetic appeal's influence on credibility applies across the three visual formats.

***H2*****:** *Visual posts with high aesthetic appeal will be perceived to be more credible than those with low aesthetic appeal.*

***RQ2*****:** *Is the positive effect of aesthetic appeal on credibility consistently observed across visual formats (photos, infographics, data visualizations)?*

***Production Quality***

Production quality refers to the extent of visual degradation or distortion to an image (Mudgal & Liu, 2022). High-quality pictures are associated with greater engagement on platforms such as *Twitter / X* and *Instagram* (Li & Xie, 2019), highlighting the importance of this visual feature. The image quality of social media posts may provide self-enhancement value increasing the overall appeal of the post separating it from the clutter of other visual posts (Li & Xie, 2019). It is important to make the distinction between production quality and aesthetic appeal, which are closely related and sometimes used interchangeably (Mudgal & Liu, 2022). Aesthetic appeal has more to do with beauty and attractiveness whereas production quality measures the degradation of the images due to factors such as noise, contrast, compression, blur, and artefacts (Mudgal & Liu, 2022). Both features impact the overall aesthetics but in different ways: aesthetic appeal focuses on the intrinsic aesthetic properties of visuals, whereas production quality is a result of extrinsic elements altering the visuals' original state.

The processing fluency framework again suggests a positive relationship between production quality of visuals and their credibility. Whittlesea et al.'s (1990) research on visual clarity, similar to how we operationalize production quality, and misattribution is relevant in this regard. In their study participants were shown a list of words in rapid succession, after which they were shown a target word whose clarity was manipulated using visual masks that introduced blurriness. Target words with greater clarity were erroneously recognized as being presented earlier, which was attributed to the metacognitive experience of familiarity due to fluency experienced by respondents. Such feelings of induced familiarity are linked with perceptions of truth, which has been studied as the 'illusory truth effect' (Begg et al., 1992).

Existing research suggests a general positive association between the quality of visuals and perceptions of trust and credibility. For example, higher quality product images in online marketplaces were associated with greater user trust in their websites (Ma et al., 2019). TV news stories presented in higher quality (resolution) videos were also rated more credible (Cummins & Chambers, 2011). Similarly, higher quality deepfake videos were perceived as more credible than lower quality deepfakes (Jin et al., 2023).

Production quality has not been systematically studied in the context of social media posts, nor in visual formats like infographics and data visualizations. We predict that high production quality visual posts will lead to greater perceived credibility, while also testing whether this effect consistently applies across the different visual formats.

> ***H3*****:** *Visuals with high production quality will be perceived to be more credible than visuals with low production quality.*
>
> ***RQ3*****:** *Is the positive effect of production quality on credibility consistently observed across different visual formats (photos, infographics, data visualizations)?*

**The Mediating Role of Processing Fluency**

As discussed above, processing fluency provides a useful framework to understand how visual features can influence the credibility of social media posts. Research suggests a clear relationship between visual features and processing fluency. Aesthetic features like visual complexity, symmetry, and color contrast influence processing fluency, such that less complex, more symmetrical, and high contrast visuals are easier to process (Reber et al., 2004; Reber & Schwarz, 1999). Similarly, visual clarity is associated with greater fluency (Whittlesea et al., 1990). Visuals having features that are easier to process elicit positive metacognitive affective reactions of fluency. Researchers have attributed this positive experience of fluency to the relative ease with which knowledge structures can be accessed, which helps in successfully interpreting the stimulus (Reber et al., 2004).

The fluency-credibility link is established through the Brunswikian model (Brunswik, 1979), as explained earlier. Fluency can act as a proximal cue to make judgements about credibility of social media posts, that is often hard to determine and hence is a distal criteria (Winkielman et al., 2012). Empirically, Hansen et al. (2008) found that participants rated statements as more credible when fluency was manipulated to be higher, by adjusting color contrast. Similarly, Parks and Toth (2006) used multiple graphic styles such as adding a spray paint and a glitch effect to manipulate fluency of text, reaching the same conclusion that processing fluency positively influences credibility judgements. Therefore, we predict that processing fluency will play a mediating role such that visuals having high aesthetic appeal and production quality, respectively, will be easier to process leading to higher credibility ratings.

***H4*****:** *Processing fluency mediates the relationships between (a) aesthetic appeal and perceived credibility and (b) production quality and perceived credibility.*

## Methods

### Study Design

We designed a preregistered, partial factorial mixed experiment embedded in a *Qualtrics* survey.[1] The three between-subjects factors were aesthetic appeal (original, low), production quality (original, low), and modality (visuals, text). The within-subjects factor for the visual conditions was visual format (photo, infographic, data visualization). Participants were randomly assigned to one of four conditions: (a) original (visuals with original aesthetic appeal and production quality); (b) low aesthetic appeal (visuals with low aesthetic appeal and original production quality); (c) low production quality (visuals with low production quality and original aesthetic appeal); and (d) text-only (control group with only text captions of social media posts without visuals). The condition with low aesthetic appeal and low production quality was not included in the experiment because this condition was not significantly different from conditions b and c in pilot tests (see Supplementary Materials Section B). As Figure 1 shows, each participant viewed 12 social media posts. To control for order effects of visual formats, we used a randomized block design that systematically varied the sequence in which participants were exposed to the three visual formats. Within each block of visual formats, the sequence of the issue contexts was further randomized.

[1] See preregistration here: https://aspredicted.org/ccp5-gnmm.pdf

**Stimuli**

The 12 stimuli used in the original condition consisted of actual social media posts published between January 2018 and April 2022, with high aesthetic appeal and high production quality visuals typical of contemporary social media. These posts were drawn from a larger dataset of visual content collected from *Twitter / X* and *Instagram* and previously rated by crowdsourced workers (Peng et al., 2025; see SM B1). The posts covered four issue domains: climate, health, the Russo-Ukrainian War (RU-War), and genetically modified organisms (GMOs). These topics were selected because they are particularly susceptible to misinformation, have the potential to generate polarization and contentious debate, and represent a diverse range of issue areas. To minimize the influence of platform or source account heuristics, each stimulus was stripped down to just its visual and accompanying text caption. The original visuals were then manipulated using *Adobe Photoshop* to create the low production quality and low aesthetic appeal conditions All stimuli can be found in Supplementary Materials (SM) Section A.

***Manipulating Aesthetic Appeal***

Aesthetic appeal is a multi-dimensional construct and there is no uniform way of manipulating it experimentally in prior research (Totti et al., 2014; Reppa & McDougall, 2022). We operationalized aesthetic appeal in terms of color, visual complexity and spatial composition.

*Color* has three aspects: hue, saturation, and brightness (Palmer et al., 2013). Hue refers to the basic color of the visual like blue, red, and yellow. Saturation refers to the intensity or purity of colors, also known as vividness or chroma. Brightness is a measure of how light or dark a color is. Prior literature establishes the important role of color in influencing aesthetic judgements, particularly in the context of website design (Cyr et al., 2010). Moreover, research suggests that preference for colors can vary based on the context they are presented in such as level of saturation, brightness and contrast (Palmer et al., 2013).

*Visual complexity* refers to the amount of variation in visuals (Peng & Jemmott, 2018). Visual complexity has been operationalized in terms of quantity of visual elements, or amount of visual details in general, with more complex designs having higher levels of information structure (King et al., 2019; Pieters et al., 2010). Research suggests that visual complexity has aesthetic qualities, with more complex designs being more engaging and preferred over less complex designs (Pieters et al., 2010).

*Spatial composition* is determined by key design principles of harmony and balance (Schloss & Palmer, 2010). Harmony refers to the audience's perception of whether two colors, or other aesthetic attributes like size, shape, or texture belong together (Alsudani & Casey, 2009).

Balance refers to the equal distribution of optical weight, or influence of size, color, and location for objects, in a visual (Alsudani & Casey, 2009). More symmetrical organization of elements leads to greater balance. In web design research, harmony and balance are considered to be important determinants of aesthetic appeal (Alsudani & Casey, 2009).

To create visuals with low aesthetic appeal, original stimuli were manipulated by altering color, visual complexity and spatial composition, wherever possible. Specifically, color aesthetics were reduced using off-beat incoherent hues or manipulating the saturation or brightness; visual complexity was manipulated by reducing visual details (number of visual elements / objects); and spatial composition was manipulated by disrupting the overall harmony or balance of the visual, such as using inconsistent font sizes or misalignment of visual elements (see stimuli examples in SM A).

***Manipulating Production Quality***

Prior research has manipulated production quality of visuals via contrast change, JPEG compression, and motion blur (Leveque et al., 2020). The current research primarily manipulates production quality by adjusting visuals' resolution, or blurriness, because it can be applied uniformly across all three visual formats. Original image resolution was reduced by adding blur filters, while ensuring that the text within the visual stimuli was still legible.

**Sample and Procedure**

Participants (N = 1200) were recruited from *Prolific*. A power analysis determined that a sample size between 1000 and 1200 participants is needed for detecting small effects at a significance level of 0.05. We employed quota sampling by gender, age, and political affiliation based on the 2020 US census. To be eligible for the study, participants had to be at least 18 years of age, fluent in English, and residing in the United States. The average survey completion time was 16.9 minutes. Each participant was paid $3.70 for completing the survey, equivalent to an average hourly rate of $15. Participants were required to complete the survey on a desktop or laptop computer instead of mobile devices to ensure proper display of all visual posts.

Respondents were asked to provide informed consent to participate in a study on social media news. After inputting their Prolific ID, they were asked to rate 12 social media posts taken from *Twitter / X* and *Instagram*. For each post, they answered questions on its perceived credibility and processing fluency. Manipulation check questions were included in the visual conditions where respondents were asked to rate the aesthetic appeal and production quality of each post (see SM Section C). Two attention checks were also included in all conditions: the first asked participants to identify the number of people shown in an image, and the second was an instructional attention check that asked participants to select "Disagree" as their answer. After rating all 12 posts, participants answered questions regarding their social media use, digital media literacy, political ideology, party affiliation, and demographics. Lastly, they were debriefed on the purpose of the study and the experimental design.

The final sample consisted of 1200 participants[2] randomly assigned to the four conditions (original condition = 299, low production quality = 303, low aesthetic appeal = 295, and text-only condition = 303). Women constituted 50.7% of the sample, men 48.4%, and non-binary participants 0.9%. In terms of racial composition, 69.5% of respondents were Caucasian, 12% were African American, 4.9% were Hispanic, 4.3% were Asian, and 8.8% reported being mixed race. 32.7% of the participants identified as Democrats, 29.1% as Republicans, and 37.8% as independents. In terms of political ideology, 44.9% of respondents ranged from liberal to extremely liberal, 21.5% as moderate, and 33.5% ranged from conservative to extremely conservative. The average age was 46.1 years old (*SD* =15.60). 33.1% of respondents had completed high school, 48.2% held a bachelor's or an associate degree, and 18.6% had a graduate degree or higher. The median household income, measured as an ordinal variable with eleven categories was $50,000 to $75,000 (*SD* = 2.31).

[2] 50 participants were excluded from the study, before being counted in the final sample, including those who attempted the survey using mobile devices (11), who did not complete the study (31), and who failed either attention check (8).

## Measures

### *Dependent Variable: Perceived Credibility*

Perceived credibility of the social media posts was measured using a four-item scale adapted from previous research (Hameleers et al., 2020; Appelman & Sundar, 2015). Participants indicated their agreement with each statement on a 7-point Likert scale: (a) "The post is credible"; (b) "The post is accurate"; (c) "The post is authentic"; (d) "The post is believable". Responses were averaged to create a composite credibility score (Cronbach's α = 0.96).

### *Mediator: Processing Fluency*

Processing fluency was measured on a four-item 7-point Likert scale (Kostyk et al., 2019): (a) "The post is easy to process"; (b) "The post is easy to understand"; (c) "The post is easy to read"; (d) "The post does not require too much time to process". Responses were averaged to create a composite processing fluency score (Cronbach's α = 0.96).

### *Covariates*

**Digital Media Literacy.** Digital media literacy was measured using the ten-item subjective media literacy scale developed by Qian et al. (2026), measuring both digital knowledge and skill. Participants indicated their level of understanding (1 = No Understanding to 5 = Full Understanding) of six internet- and computer-related terms: PDF, Spyware, Phishing, Hashtag, Chatbot, and Algorithm. To assess digital skills, participants indicated their agreement with four statements such as “I know how to solve my own technical problems on digital devices and platforms” and “I can learn new technologies easily.” A composite score averaging the ten items (Cronbach’s $\alpha = 0.91$) was used as a covariate in the analysis.

**Political Conservatism.** Political ideology was measured by asking participants to place themselves on a 7-point scale ranging from “Extremely Liberal” to “Extremely Conservative.” We also asked the participants to indicate their party affiliation (Democrat, Republican, Independent, or Other). Only political ideology was included as a covariate in the analyses (Mean = 3.72, Median = 4: “Moderate / Center”, SD = 1.86).

**Demographics.** Age, gender, and education were included in the analyses. Age was measured as a continuous variable. Gender, measured with three categories, was converted into a binary variable for analyses (female =1, male or non-binary = 0). Education was measured as an ordinal variable, ranging from less than high school to doctorate degree or higher (Median = 5: “Bachelor’s Degree”, *SD* = 1.41), and was converted into a binary variable with associate degree or lower as the reference category. All descriptives and correlations are shown in Table D1.

**Analytical Strategy**

To test H1-H3 and RQ1-RQ3, we fitted mixed-effects linear regression models predicting perceived credibility by including random effects (intercept) for participants for variance partitioning as each participant rated twelve posts in total. An intercept-only model was run to calculate the Intraclass Correlation Coefficient, which was 0.308, suggesting that approximately 31% of total variance in credibility ratings was attributable to individual-level differences between respondents, warranting the use of these models. To test H1 and RQ1, visual formats (visuals = 1, text = 0) as well as dummy variables of the photo, infographic, and data visualization formats were used as main predictors. To test H2, H3, RQ2, RQ3, and for the mediation analyses, low aesthetic appeal and low production quality conditions were compared against the original condition. We estimated each model both with and without covariates, including age, political ideology, gender, education, and digital media literacy. The results section reports findings from models with covariates, and the models without covariates can be found in SM Table E1.

To test the mediation hypotheses (H4a-b), we ran Bayesian multilevel mediation models. Estimation was based on four Markov Chain Monte Carlo chains with minimum 8000 iterations per chain and weakly informative priors. The posterior distributions of the direct, indirect, and total effects were used to evaluate the mediation. Posterior summaries are presented in the results section including posterior median (M) values and 90 percent credible intervals (CrI). The mediation was considered significant if the interval did not contain zero (Enders et al., 2013). Additional post hoc mediation analyses were run including mediation of processing fluency between visual features and credibility for each visual format separately.

## Results

### Superior Visual Effects and Differences across Visual Formats

H1 predicted that social media posts containing visuals would be perceived as more credible than text-only posts. The analyses revealed that visual social media posts indeed were perceived as more credible than text-only posts ($B = 0.29$, $SE = 0.06$, $p < .001$), supporting H1. However, this superiority effect of visuals over text did not apply consistently across the three visual formats (RQ1), as shown in Figure 2. Posts with photos were rated as more credible than their text-only counterparts ($B = 0.15$, $SE = 0.06$, $p = .017$), and a similar effect was found for infographics ($B = 0.61$, $SE = 0.07$, $p < .001$). However, data visualizations were not rated as significantly more credible than text-only posts (*B* = 0.10, *SE* = 0.07, $p = .117$; see Models 1 to 4 in SM Table E2 for detailed regression results).

**Visual Features and Credibility**

H2 predicted that visual social media posts with high aesthetic appeal will be perceived as more credible than those with low aesthetic appeal. The results of the mixed effects models supported H2 as original visual posts were rated as significantly more credible than those in the low aesthetic appeal condition (*B* = 0.17, *SE* = 0.06, $p = .009$). The positive effect of aesthetic appeal on credibility did not apply uniformly across formats (RQ2); aesthetic appeal was associated with increased credibility perceptions of social media posts with infographics (*B* = 0.25, *SE* = 0.07, $p < .001$) and data visualizations (*B* = 0.24, *SE* = 0.08, $p = .002$), but had no effect for photos (*B* = 0.02, *SE* = 0.07, $p = .838$) (see Models 5 to 8 in SM Table E2).

H3 predicted that posts with high production quality visuals will be perceived as more credible than those with low production quality visuals. Production quality did not influence perceived credibility for visuals overall (*B* = 0.05, *SE* = 0.06, $p = .434$). H3 was not supported. This was consistent across all visual formats (RQ3).

Among the covariates, respondents' digital media literacy and political ideology were generally significant predictors of credibility. Across all models, except for the one testing aesthetic appeal of photos (Model 6 in SM Table E2), digital media literacy positively predicted credibility. By contrast, political conservatism in general negatively predicted credibility ratings (Models 2, 3, 5, 6, and 7 in SM Table E2).

**Mediation Effects of Processing Fluency**

H4a predicted that processing fluency will mediate the relationship between aesthetic appeal and perceived credibility of visual posts. Results from the Bayesian multilevel mediation models (SM Table F1) showed that for aesthetic appeal, the indirect effect on credibility via processing fluency was small but significant (*M* = 0.04, 90% CrI [0.008, 0.071]). The direct (*M* = 0.13, 90% CrI [0.038, 0.217]) and total (*M* = 0.17, 90% CrI [0.074, 0.261]) effects of aesthetic appeal on credibility were also significant. These results indicate that processing fluency partially mediates the relationship between aesthetic appeal and perceived credibility of social media posts. Therefore, H4a was supported.

H4b predicted that processing fluency would mediate the relationship between production quality and perceived credibility of visual posts. The indirect effect on credibility through processing fluency was significant (*M* = 0.05, 90% CrI [0.017, 0.075]). However, the direct (*M* = 0.01, 90% CrI [–0.083, 0.094]) and total (*M* = 0.05, 90% CrI [–0.042, 0.145]) effects were not significant, providing inconclusive evidence of mediation. Therefore, H4b was not supported.

**Post Hoc Analyses**

To further explore whether processing fluency explains the superiority effect of visuals over text on credibility ratings, we ran mediation analyses for all visuals and for individual formats (see SM Table F2). For overall visuals, processing fluency did not mediate the relationship between the visual modality and credibility, as the indirect effect was not significant (*M* = – 0.03, 90% CrI [–0.057, 0.001]). Mediation analyses for individual visual formats, however, revealed that processing fluency fully mediated the relationship between photos (versus text) and credibility (Indirect effect: *M* = 0.24, 90% CrI [0.184, 0.286]; Direct effect: *M* = –0.08, 90% CrI [–0.170, 0.006]; Total effect: *M* = 0.15, 90% CrI [0.053, 0.255]), and partially mediated the relationship between infographics (versus text) and credibility (Direct effect: *M* = 0.36, 90% CrI [0.266, 0.445]; Indirect effect: *M* = 0.26, 90% CrI [0.209, 0.304]; Total effect: M = 0.61, 90% CrI [0.512, 0.712]). The results suggest that posts with photos and infographics are more easily processed than the ones relying on text alone, which at least partially explains why such posts are rated as more credible than text-only posts.

We also further dissected the mediation effect of processing fluency between visual features and credibility ratings (H4a-b) by running mediation models for individual formats (See SM Table F1). For infographics, processing fluency fully mediated the relationship between aesthetic appeal and perceived credibility of posts (Indirect effect: $M = 0.21$, 90% CrI [0.166, 0.260]; Direct effect: $M = 0.04$, 90% CrI [–0.066, 0.138]; Total effect: $M = 0.25$, 90% CrI [0.139, 0.359]). For data visualizations, the indirect effect was negative ($M = –0.05$, 90% CrI [–0.095, –0.011]) and dampened the direct effect ($M = 0.29$, 90% CrI [0.184, 0.398]), reducing the overall effect ($M = 0.24$, 90% CrI [0.122, 0.355]). For photos, the mediation effect was not found. These results suggest that processing fluency mediates the effect of aesthetic appeal on credibility most strongly for infographics, and in fact works in the opposite direction for data visualizations. Mediation analyses between production quality and perceived credibility for all three formats in the post hoc analyses were not significant (See SM Table F1).

## Discussion

The exponential growth of visual media challenges credibility theories originally built around text-based communication. Drawing on processing fluency theory, this study examined *whether* visuals enhance perceived credibility and *how* specific visual formats and features and cognitive mechanisms help explain this effect. Using a preregistered online experiment with 1,200 US participants, we compared three prevalent visual formats—photos, infographics, and data visualizations—to text-only posts in terms of credibility, and further assessed the roles of aesthetic appeal, production quality, and processing fluency. Overall, we found that social media posts containing visuals were perceived as more credible than text-only posts, but this advantage was not uniform across formats. The credibility premium only applied to infographics and photos, whereas data visualizations did not significantly outperform text-only posts. Aesthetic appeal increased perceived credibility—particularly for infographics and data visualizations—and this relationship was partially mediated by processing fluency. In contrast, production quality did not significantly influence credibility of social media posts. Together, these findings suggest that while visual social media posts are generally perceived as more credible than text-only posts, their credibility advantage varies by visual format and depends on how aesthetically appealing and easy to process they are.

**Visuals' Credibility Premium Varies by Format and Aesthetic Appeal**

A central finding is that the credibility premium of visuals is not uniform across formats. This is consistent with previous research showing that photos are perceived as more credible than text (Newman & Schwarz, 2024; Smelter & Calvillo, 2020), and extends this finding further to infographics. This suggests that visuals combining structured information with graphic elements may provide both informational richness and perceptual accessibility, thereby enhancing credibility judgments. By contrast, data visualizations did not yield a similar advantage over text-only posts, one possible explanation of which is that data visualizations contain dense information and hence are less fluently processed. This pattern suggests that visuals do not inherently increase credibility; instead, their persuasive value appears contingent on how effectively they organize and communicate information. Together, these findings highlight the importance of differentiating among visual formats in credibility research rather than treating visuals as a homogeneous category.

The role of aesthetic appeal further clarifies these format differences. Consistent with prior research (Reber et al., 2004), visually appealing posts were generally perceived as more credible, especially in the case of infographics and data visualizations. Mediation analyses showed that processing fluency partially explains this relationship: aesthetically appealing visuals are easier to process, and this subjective ease translates into higher credibility judgments. Notably, this mediating effect was strongest for infographics, suggesting that aesthetic features including color, visual complexity, and spatial composition may be particularly consequential when information is presented in structured, explanatory formats. For data visualizations, however, the mediation pattern was more nuanced. Although aesthetic appeal increased credibility directly, processing fluency exerted a negative indirect effect, suggesting that even aesthetically appealing data visualizations may impose cognitive demands due to their density or analytical nature. As a result, perceptual ease does not uniformly translate into higher credibility and instead, the relationship between fluency and credibility may depend on the interplay between visual appeal and informational complexity.

Contrary to our expectation, production quality did not significantly affect credibility across formats, nor did it demonstrate clear mediation through fluency. This aligns with some previous literature suggesting that people are bad at comparative judgements of image quality (Cummins & Chambers, 2011). This null effect suggests that audiences may not treat technical refinement—such as resolution—as a primary credibility cue in social media contexts. One explanation is that widespread exposure to user-generated and lower-quality visuals has normalized variability in production standards, reducing their diagnostic value.

## Implications

Theoretically, this research extends the literature on visual credibility in three ways. First, it extends credibility research beyond photographs to two understudied visual formats, infographics and data visualizations, and demonstrates visuals' credibility advantage over text does not apply evenly across different visual formats. These results imply that treating visuals as a homogeneous category obscures meaningful variation. Second, it foregrounds visual features—particularly aesthetic appeal and production quality—as primary and independently-operationalized dimensions of analysis rather than treating them as part of the visual modality bundle. By clarifying and operationalizing these constructs within the context of contemporary social media, the study contributes conceptual refinement to areas that have previously lacked scholarly consensus. The findings corroborate prior evidence that aesthetics positively influence credibility (Robins & Holmes, 2008), while extending this insight to ecologically valid stimuli that resemble real-world social media posts across multiple visual formats. In doing so, it strengthens the external validity of visual credibility research, which has often relied on simplified stimuli such as text or abstract shapes (Reber & Schwarz, 1999; Reber et al., 1998; Whittlesea et al., 1990). Third, by testing processing fluency as a mediating mechanism, the research demonstrates that the ease with which users process aesthetically appealing visuals, particularly infographics and data visualizations, partially explains their credibility advantage. It thereby offers a more precise theoretical account of how visual features shape credibility judgments in digital environments, with clear boundary conditions, while also extending previous fluency research which primarily relied on manipulation of simple text stimuli (Reber & Schwarz, 1999; Parks & Toth, 2006). As communication research engages an increasingly visual public sphere, fluency-based mechanisms offer a portable theoretical bridge across subfields whose stimuli, audiences, and stakes differ but whose underlying credibility processes share a common ground.

Practically, the results suggest that communicators seeking to enhance perceived credibility should prioritize visual aesthetics. Increasing technical production quality alone may not yield credibility gains. For complex formats such as data visualizations, reducing informational density and enhancing interpretability may be more important than improving surface polish. Meanwhile, if aesthetic appeal increases credibility through fluency, visually appealing misinformation may benefit from this heuristic pathway. As advances in artificial intelligence make it easier to produce polished and aesthetically compelling visuals, misleading content may gain credibility advantages over accurate content with limited visual appeal. This finding also suggests that media organizations and fact-checkers could prioritize visually appealing content for their fact-checking efforts. More broadly, the results underscore the importance of enhancing media literacy efforts to address fluency-based biases and may help users become more reflective about the influence of visual design on their credibility evaluations in increasingly visual digital spaces.

**Limitations & Future Research**

This study has a few limitations that warrant further research. First, due to the experimental nature of the study, participants were only exposed to social media posts with very limited context, including a text caption and visual (if applicable). They were unable to click through URLs, nor were they allowed to interact with other elements typically available on social media platforms (e.g., account information, comments section, etc). The lack of context may have influenced their appraisal of these posts' credibility. Second, participants were not asked about their pre-existing attitudes towards the four contentious issues examined in the study. This can be a potential limitation as previous research has found respondents' pre-existing attitudes to be a significant predictor of image credibility evaluations (Shen et al., 2019). Lastly, due to the inherent differences in the visual stimuli across three formats, our experimental manipulation of aesthetic appeal and production quality might not be uniform across all stimuli.

This study provides many avenues for future research. Future work can test specific aesthetic dimensions, such as color, visual complexity, and balance / harmony, separately to see which is more instrumental in influencing credibility perceptions. Future research can also incorporate additional outcome variables such as engagement intention and discernment of social media posts. Another potential direction is to study the visual features of videos and memes, which are also popular visual formats not included in the current study. Lastly, with the advancement of LLMs, it might become more feasible to study visual features, even those that are more subjective in nature, of larger samples of social media posts. This can be done to complement experimental methods, or in conjunction with human annotation to validate results.

## Data Availability Statement

The data and code to replicate the analysis results can be found at: https://osf.io/k9tg4/overview?view_only=085ff22b3bef4b35abe742d8ca8e1ac2.

**Figure 1.** Experimental Design and Survey Flow

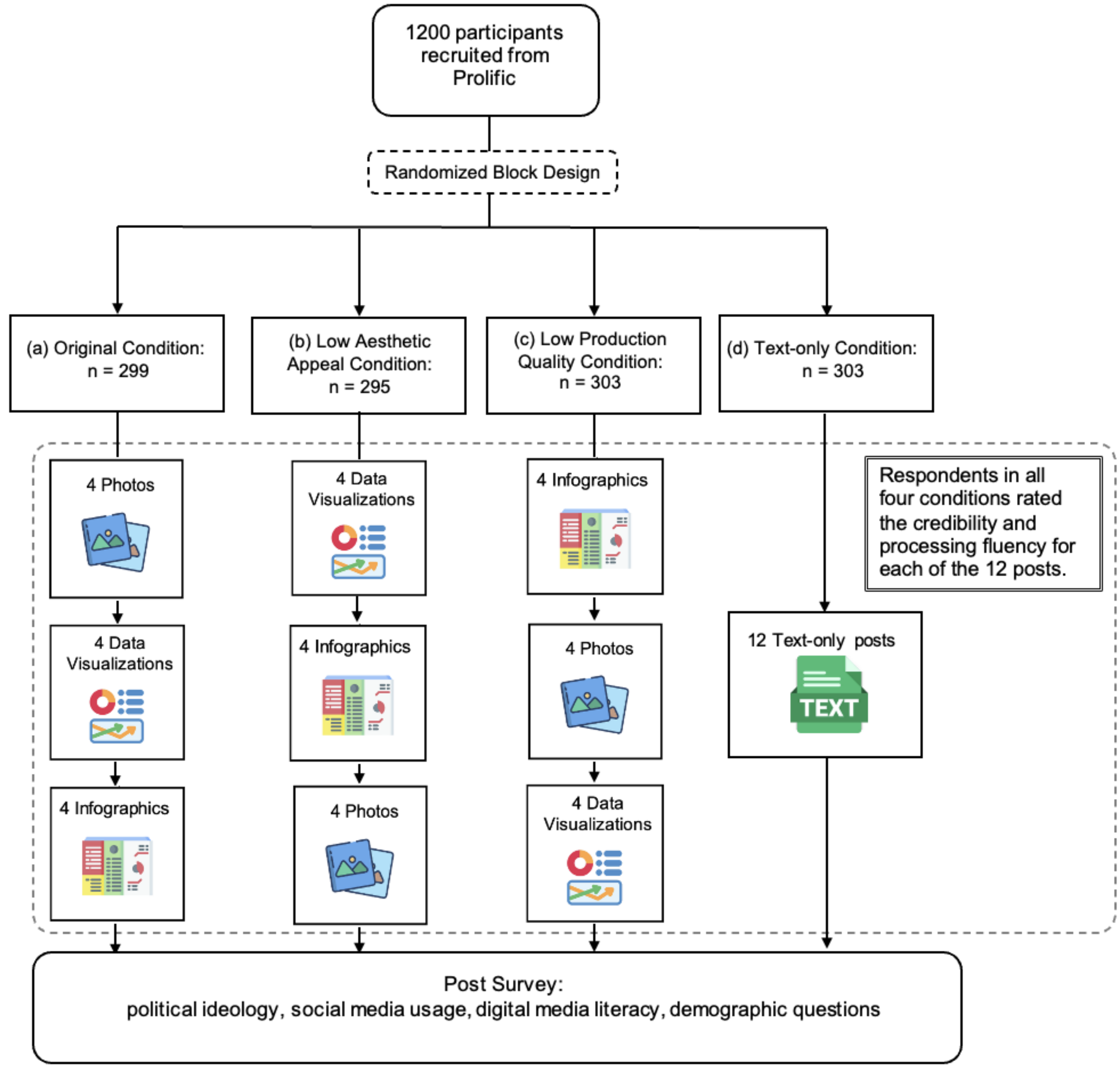

**Figure 2**. Visual Formats and Features Predicting Perceived Credibility of Social Media Posts

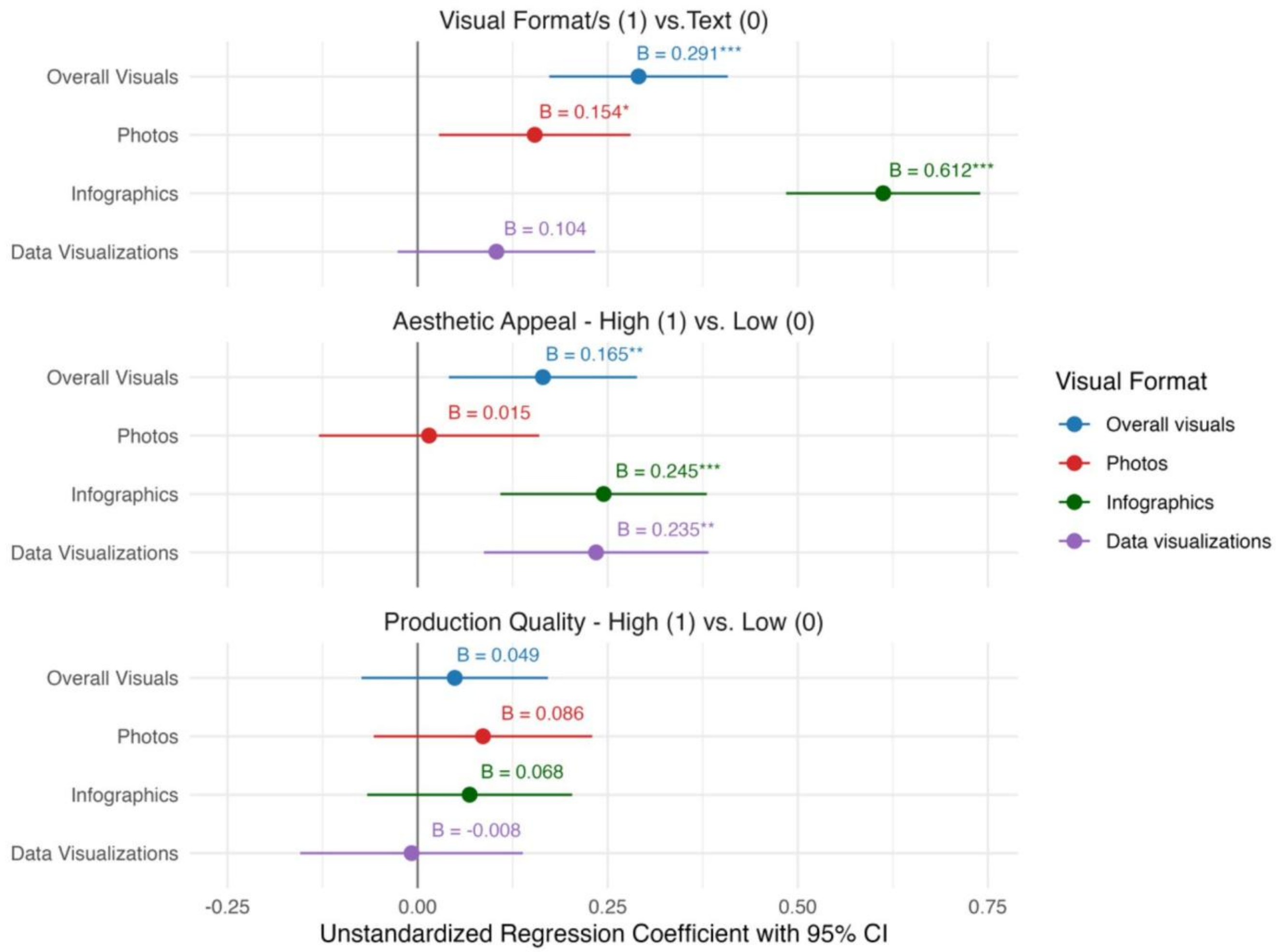

*Notes:* Figure shows unstandardized regression coefficients. Dummy variables for visuals (ref. category = text-only posts) and visual features (ref. category = low aesthetic appeal / low production quality) were used in the mixed-effects models with covariates being age, gender, digital media literacy, education, and political conservatism. $^{*}p < .05$, $^{**}p < .01$, $^{***}p < .001$.

# Supplementary Materials (SM)

**A. Stimuli**

**A1. Photos**

A1.1. Original Condition

Climate

Around the world, poor and vulnerable communities are facing multiple crises at once. Sadly, they have contributed least to the #climatecrisis but, are being hit the hardest. https://t.co/9mxf4t6J5T

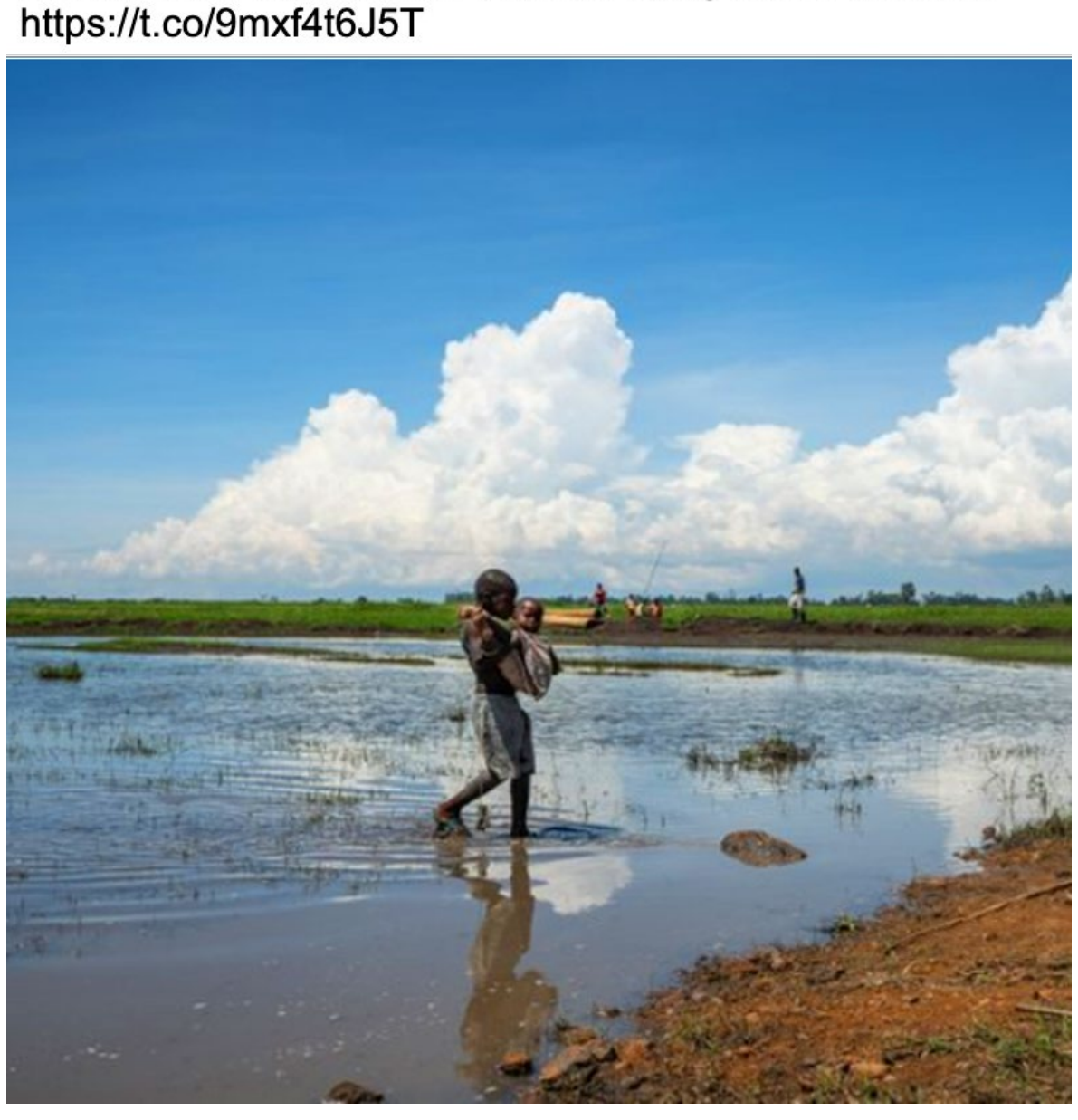

Health

Vaccines💉 are the world's safest way to protect children from serious diseases such as polio, measles and smallpox. They help children grow up healthy and happy. #VaccinesWork #ForEveryChild @gavi @Sida @WHO @SanteTchad @OnuTchad https://t.co/Km3iBmkEWc

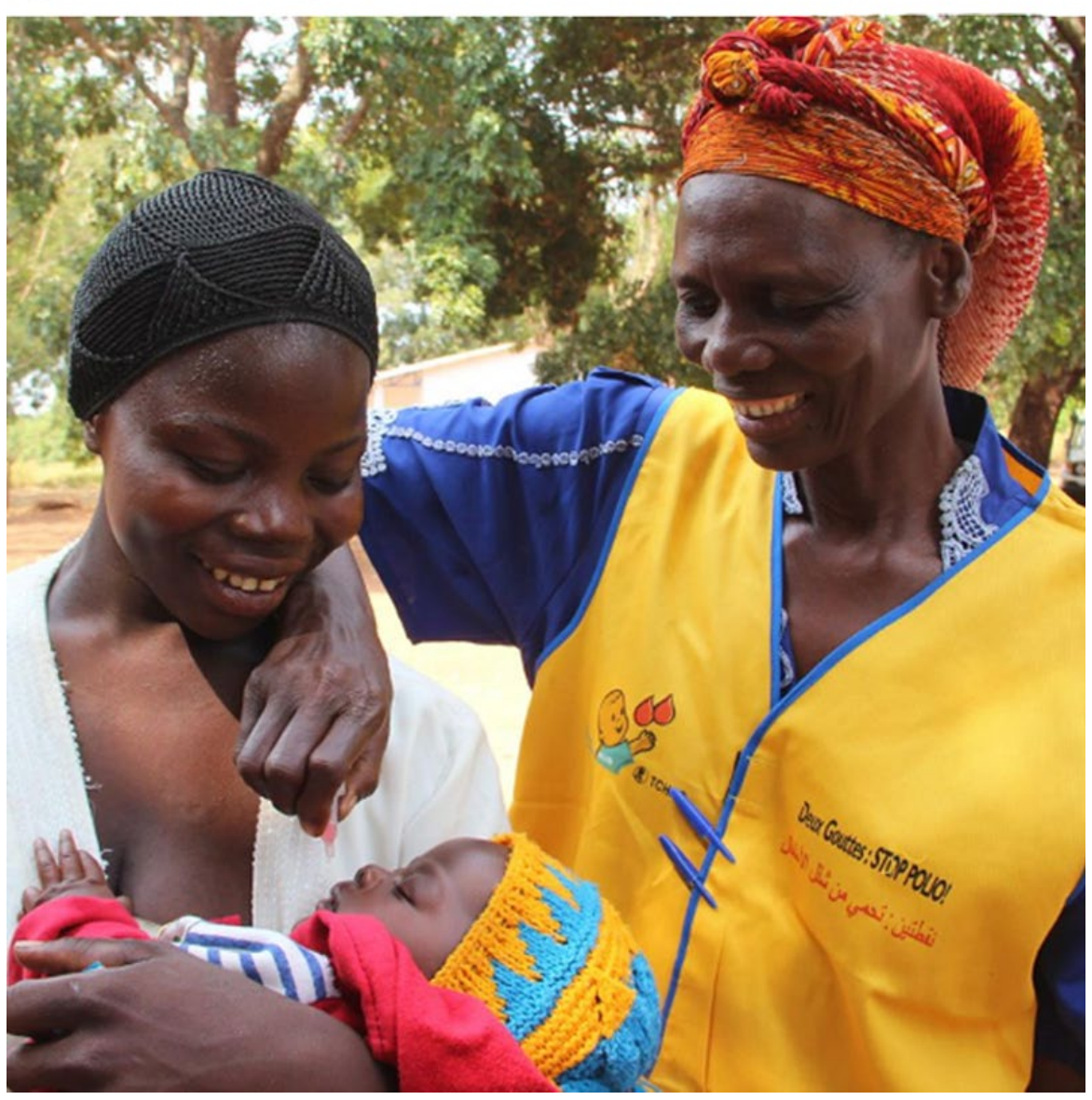

Russo-Ukrainian War

GMO

Russia said its #missile cruiser #Moskva has been seriously damaged after an ammunition explosion, Interfax has reported. The #crew of the Moskva has been evacuated and the cause of the fire that prompted the explosion is being investigated. 🇷🇺🇺🇦 #RussianUkrainianWar https://t.co/cFOCIHK4tQ

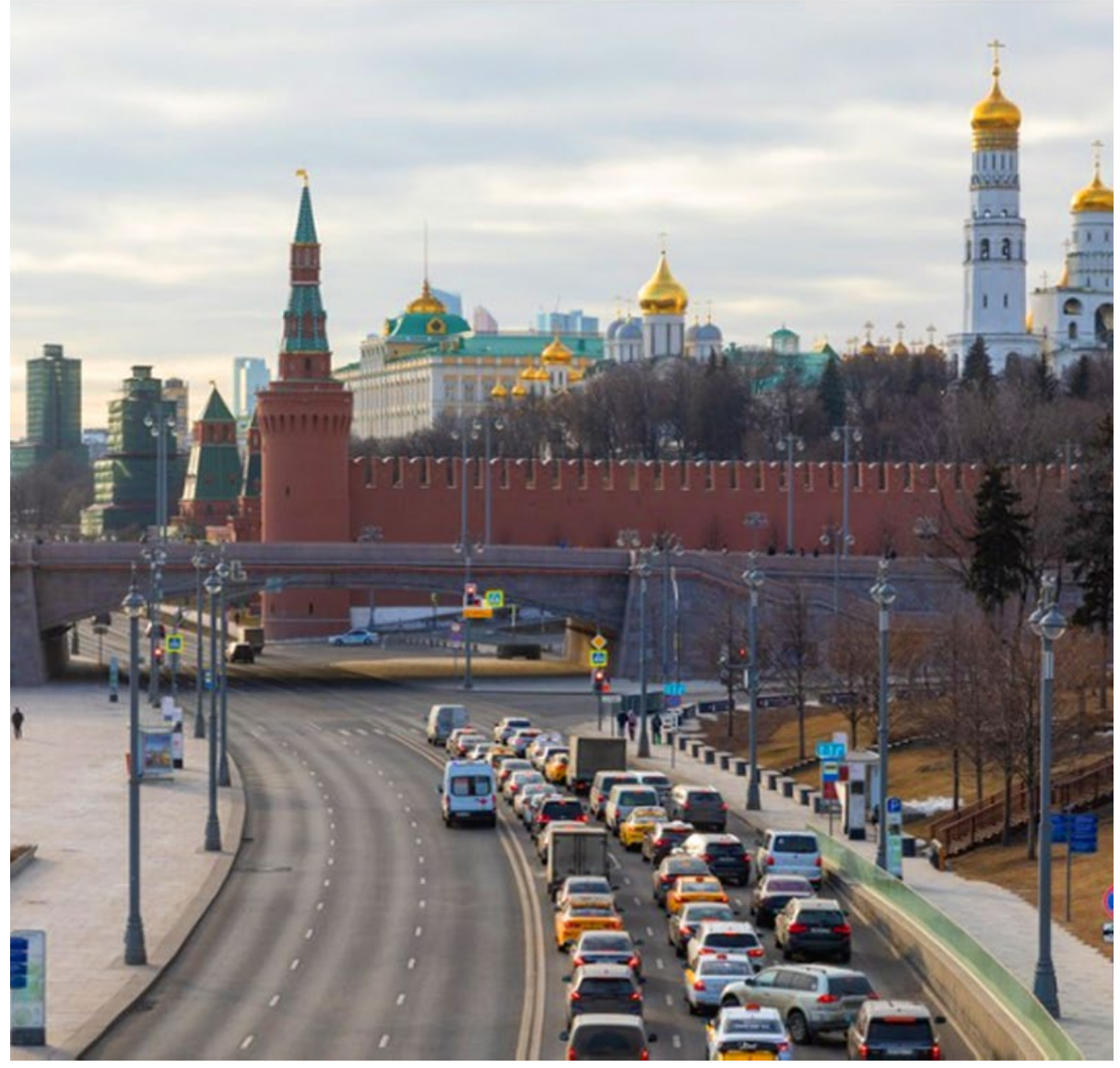

Just how bad are GMO's? Read this article for what experts say about GMO's and if they're safe to consume. https://t.co/WAgyBsMCl7 #gmo #nutrition #health https://t.co/DilcKUJBCF

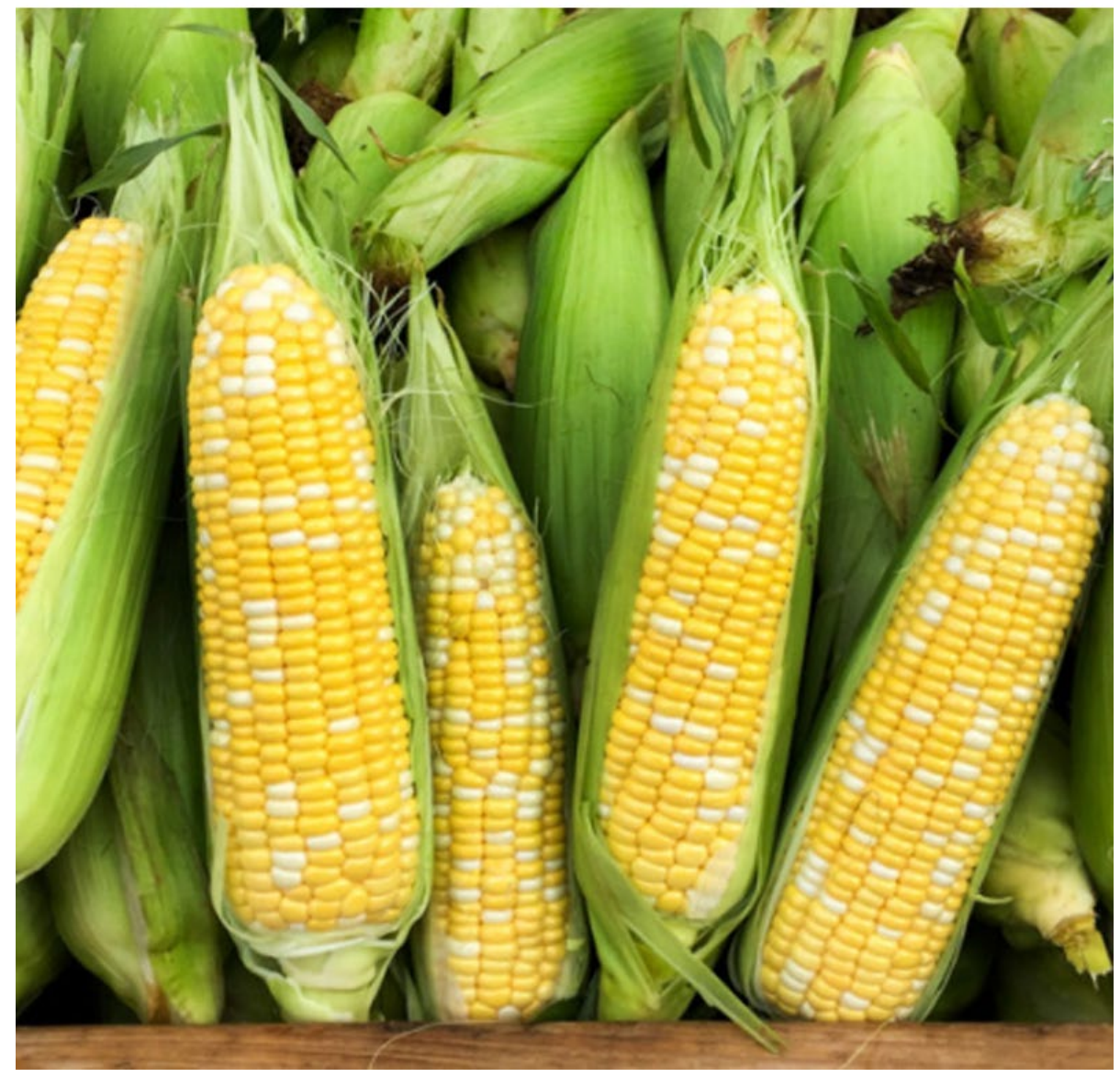

## A1.2. Low Aesthetic Appeal Condition

### Climate

Around the world, poor and vulnerable communities are facing multiple crises at once. Sadly, they have contributed least to the #climatecrisis but, are being hit the hardest. https://t.co/9mxf4t6J5T

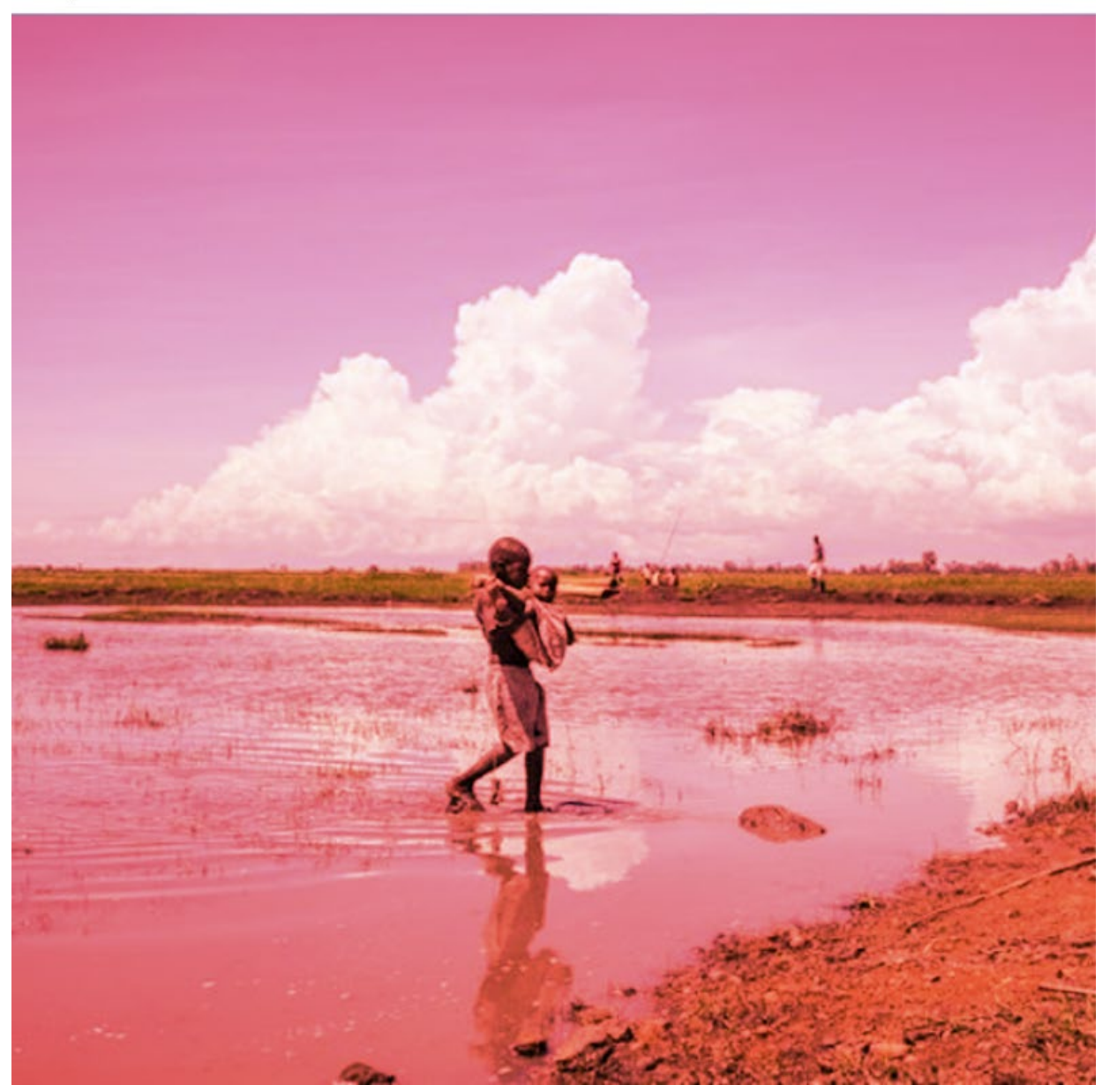

### Health

Vaccines💉 are the world's safest way to protect children from serious diseases such as polio, measles and smallpox. They help children grow up healthy and happy. #VaccinesWork #ForEveryChild @gavi @Sida @WHO @SanteTchad @OnuTchad https://t.co/Km3iBmkEWc

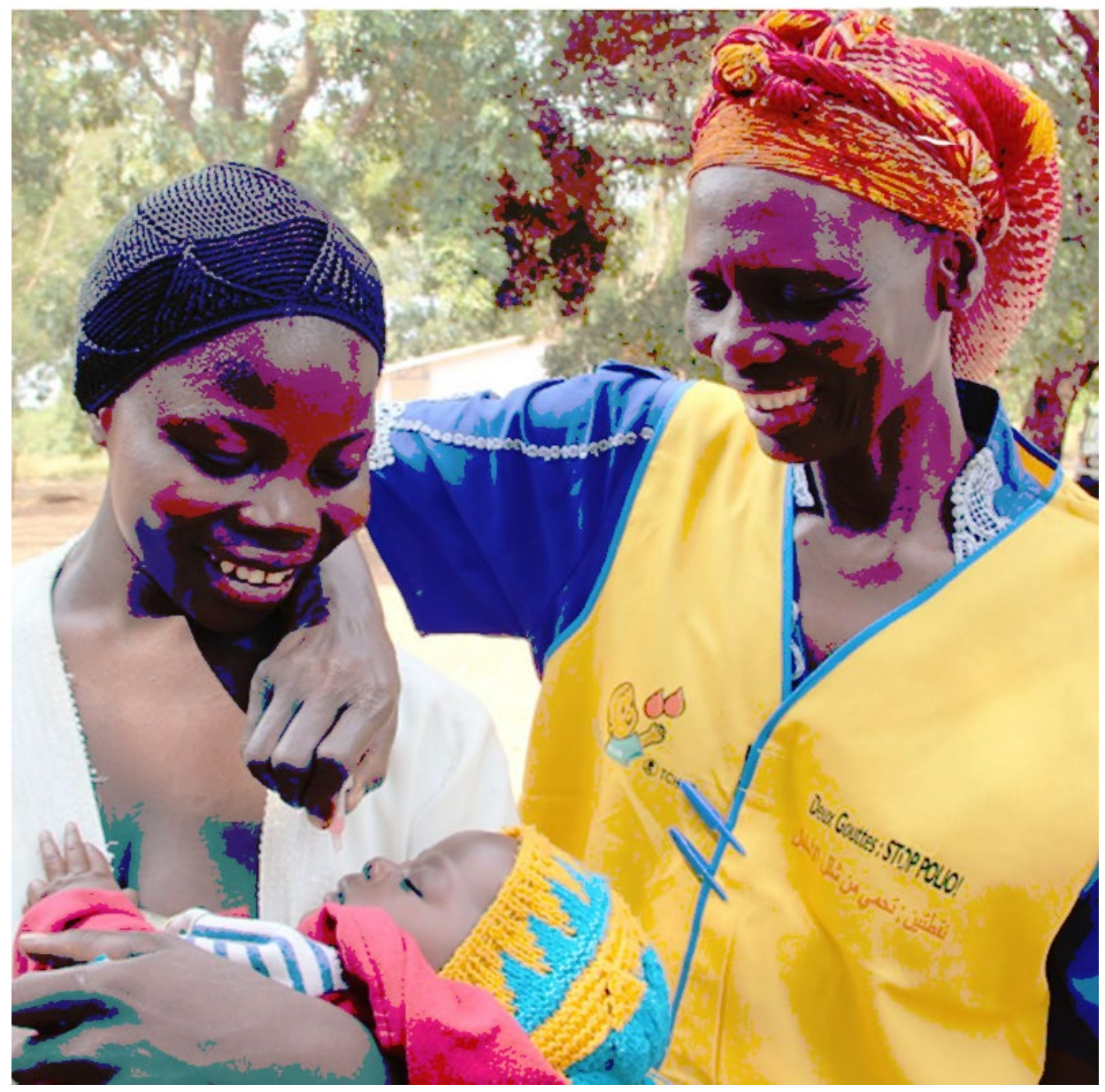

Russo-Ukrainian War

Russia said its #missile cruiser #Moskva has been seriously damaged after an ammunition explosion, Interfax has reported. The #crew of the Moskva has been evacuated and the cause of the fire that prompted the explosion is being investigated. 🇷🇺🇺🇦
#RussianUkrainianWar https://t.co/cFOCIHK4tQ

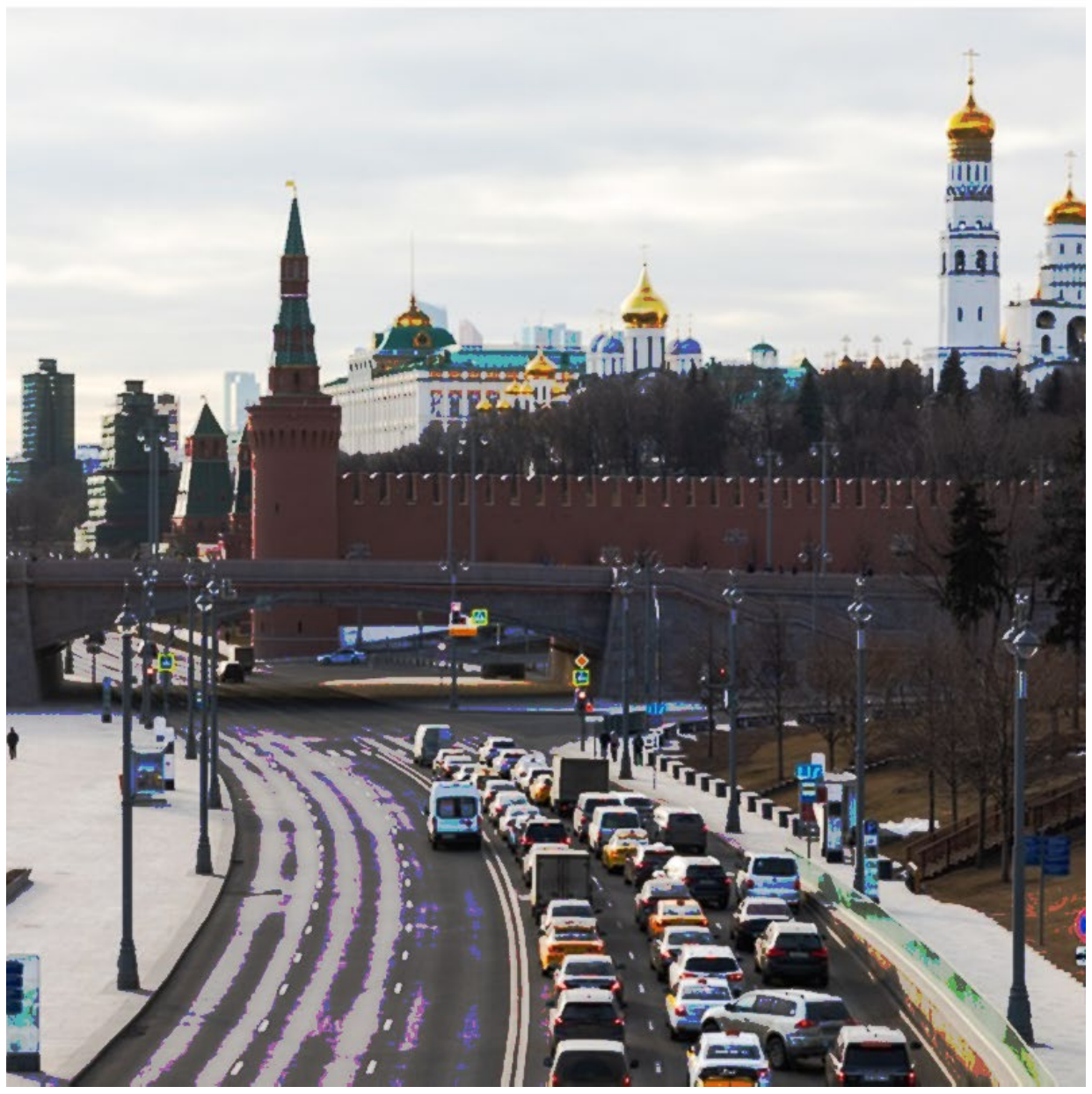

GMO

Just how bad are GMO's? Read this article for what experts say about GMO's and if they're safe to consume. https://t.co/WAgyBsMCl7 #gmo #nutrition #health https://t.co/DilcKUJBCF

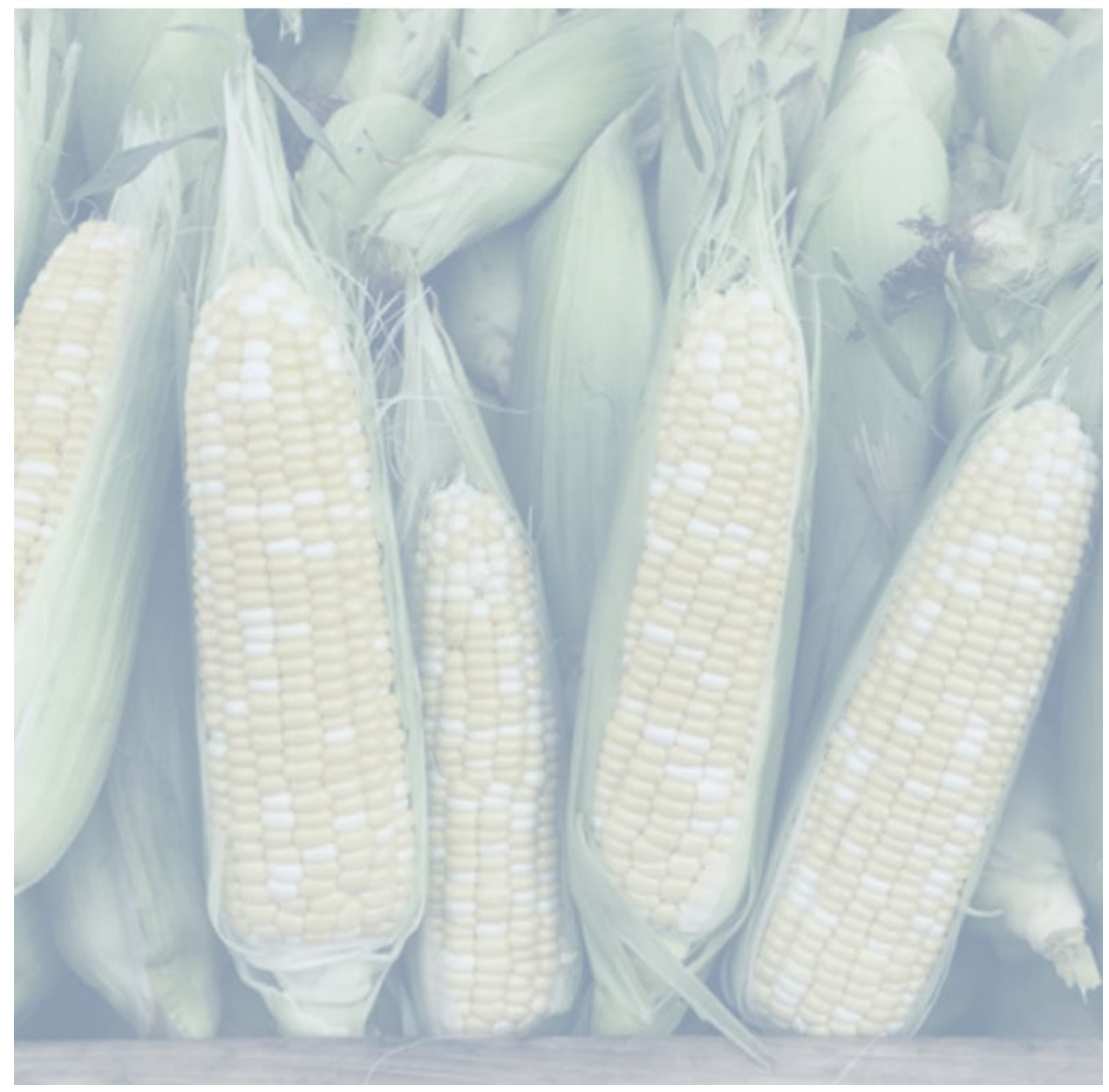

A1.3. Low Production Quality Condition

Climate

Around the world, poor and vulnerable communities are facing multiple crises at once. Sadly, they have contributed least to the #climatecrisis but, are being hit the hardest. https://t.co/9mxf4t6J5T

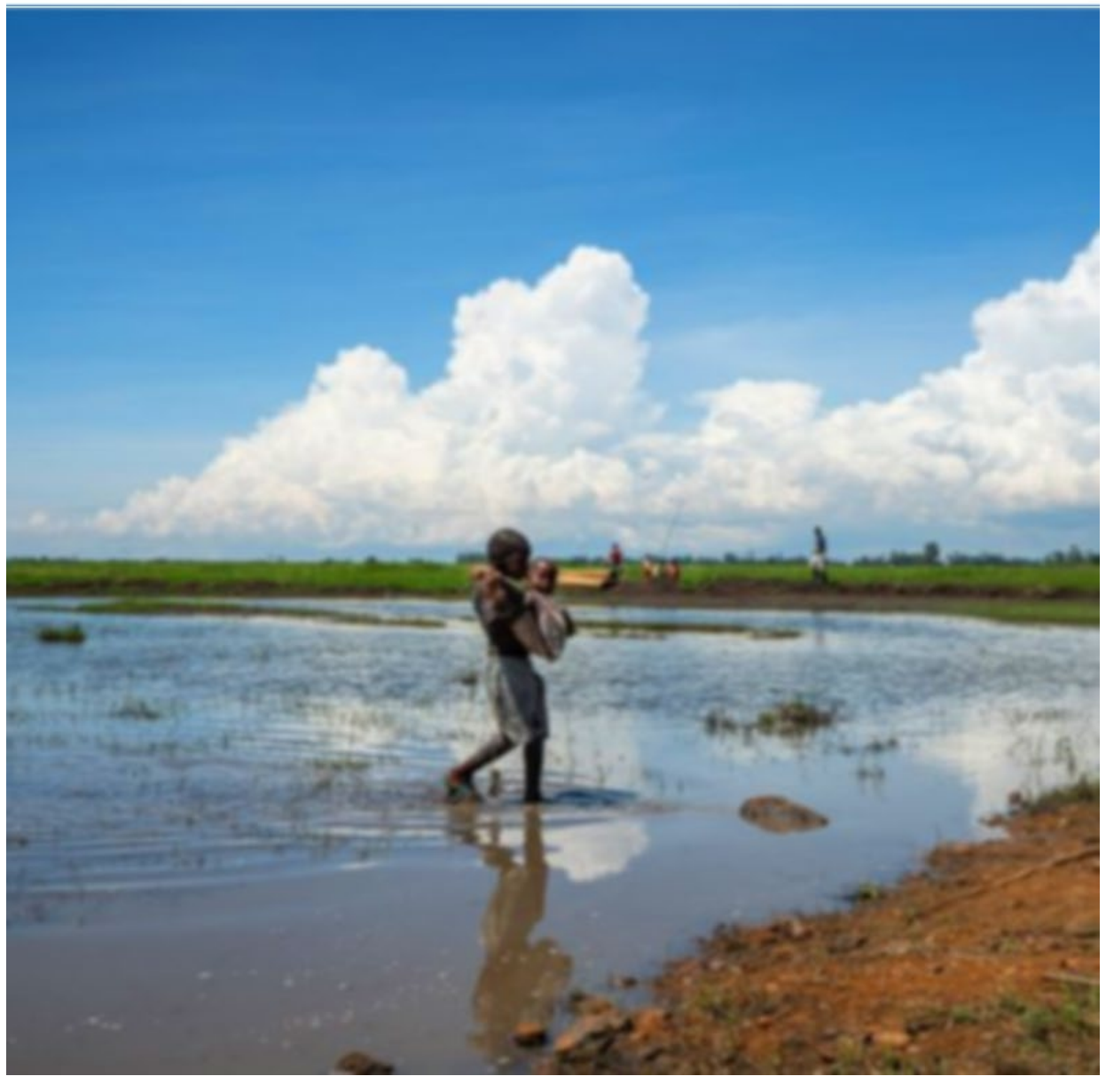

Health

Vaccines💉 are the world's safest way to protect children from serious diseases such as polio, measles and smallpox. They help children grow up healthy and happy. #VaccinesWork #ForEveryChild @gavi @Sida @WHO @SanteTchad @OnuTchad https://t.co/Km3iBmkEWc

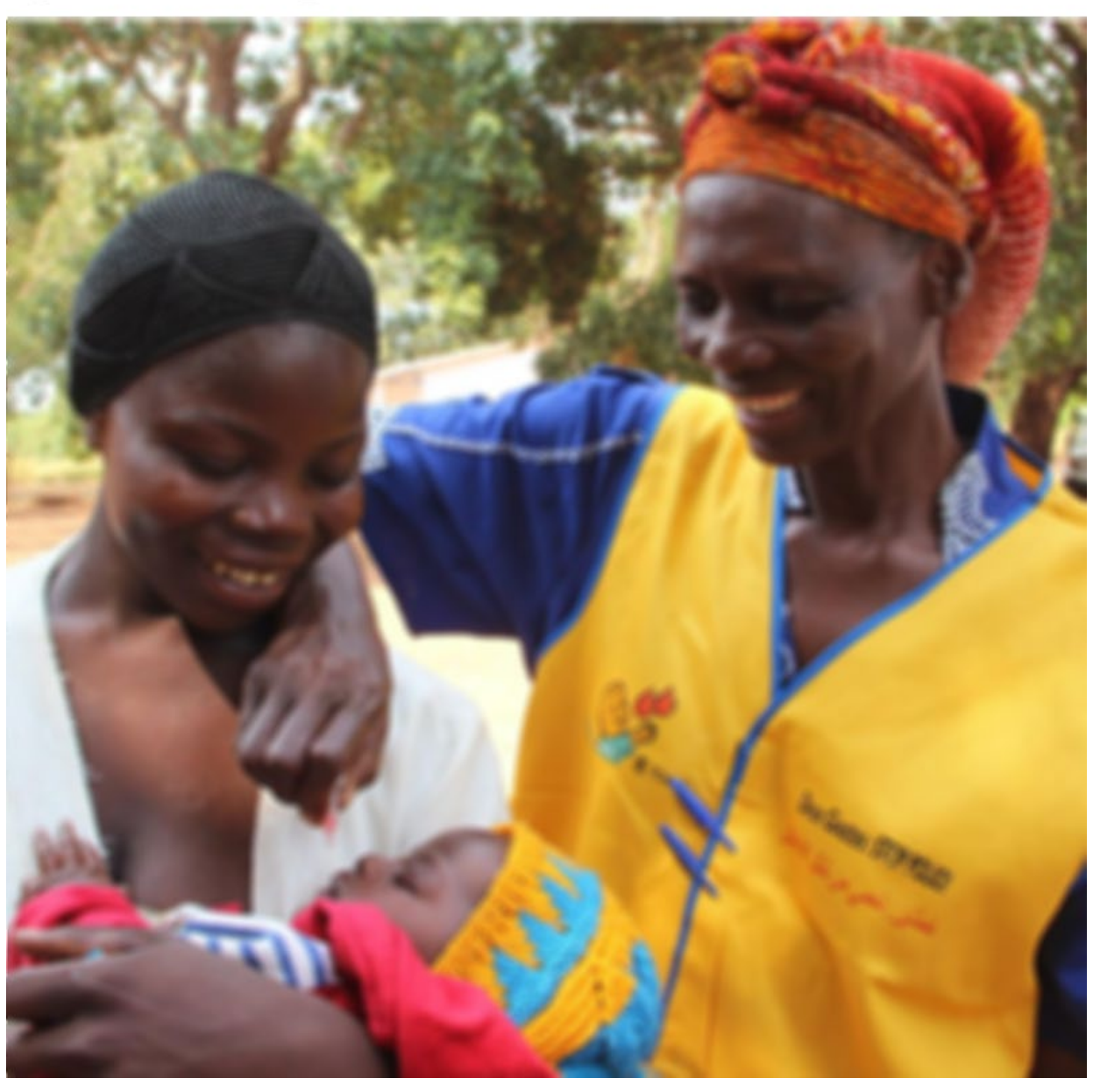

Russo-Ukrainian War

GMO

Russia said its #missile cruiser #Moskva has been seriously damaged after an ammunition explosion, Interfax has reported. The #crew of the Moskva has been evacuated and the cause of the fire that prompted the explosion is being investigated. 🇷🇺🇺🇦 #RussianUkrainianWar https://t.co/cFOCIHK4tQ

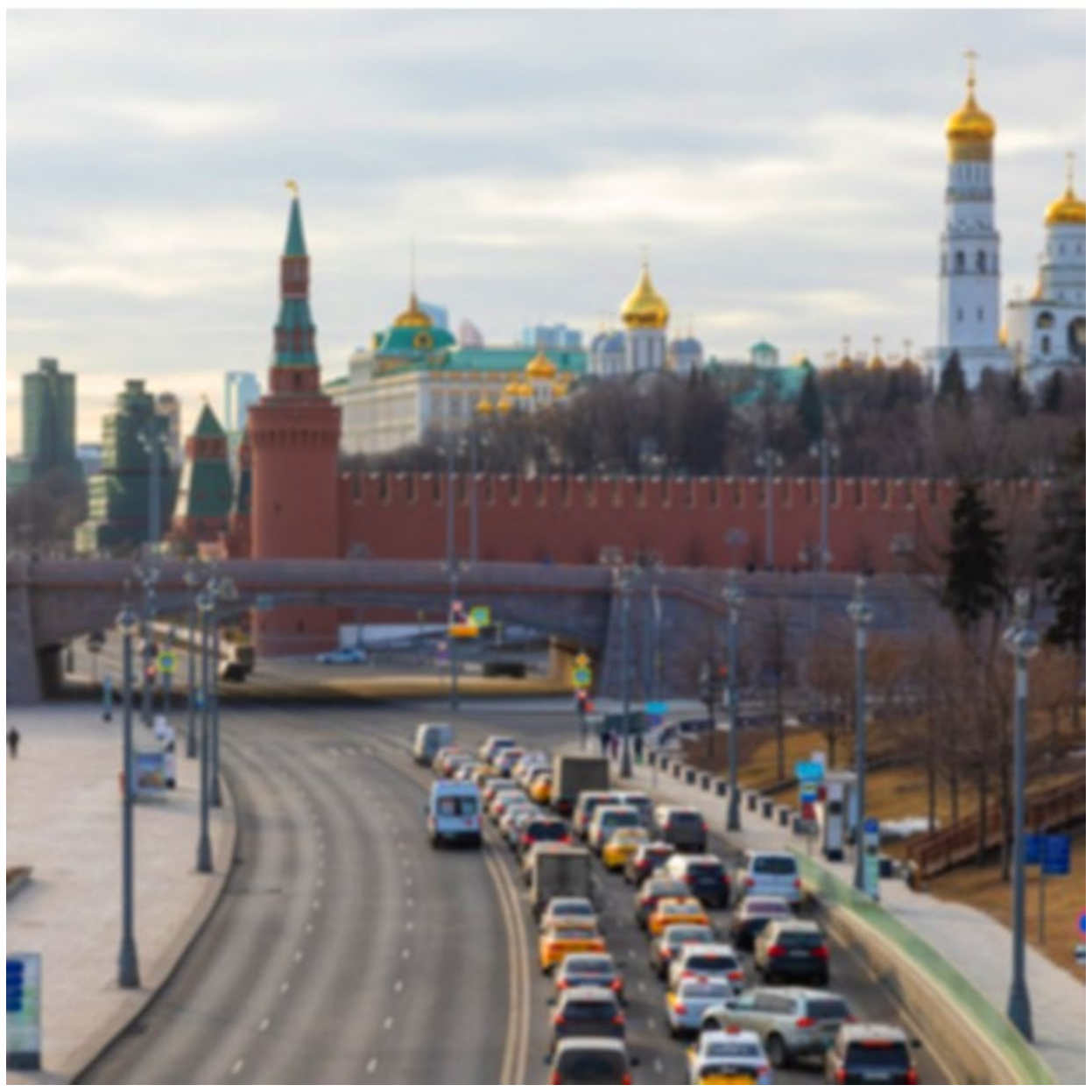

Just how bad are GMO's? Read this article for what experts say about GMO's and if they're safe to consume. https://t.co/WAgyBsMCl7 #gmo #nutrition #health https://t.co/DilcKUJBCF

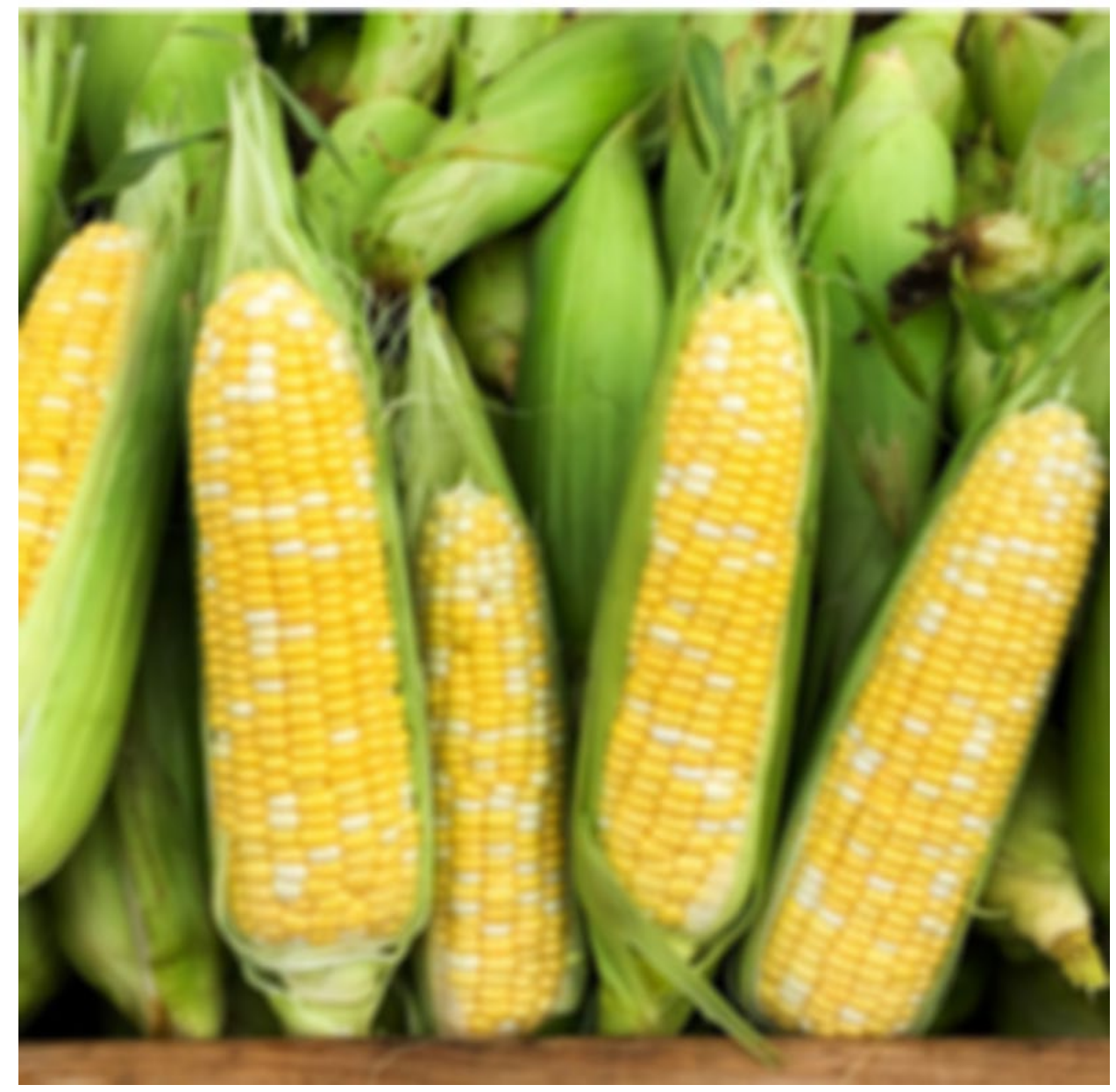

## A2. Infographics

### A2.1. Original Condition

Climate

Pay attention people. Heat is the deadliest form of weather. Be safe!!! #HEATWAVE https://t.co/sUQ0WdfSmk

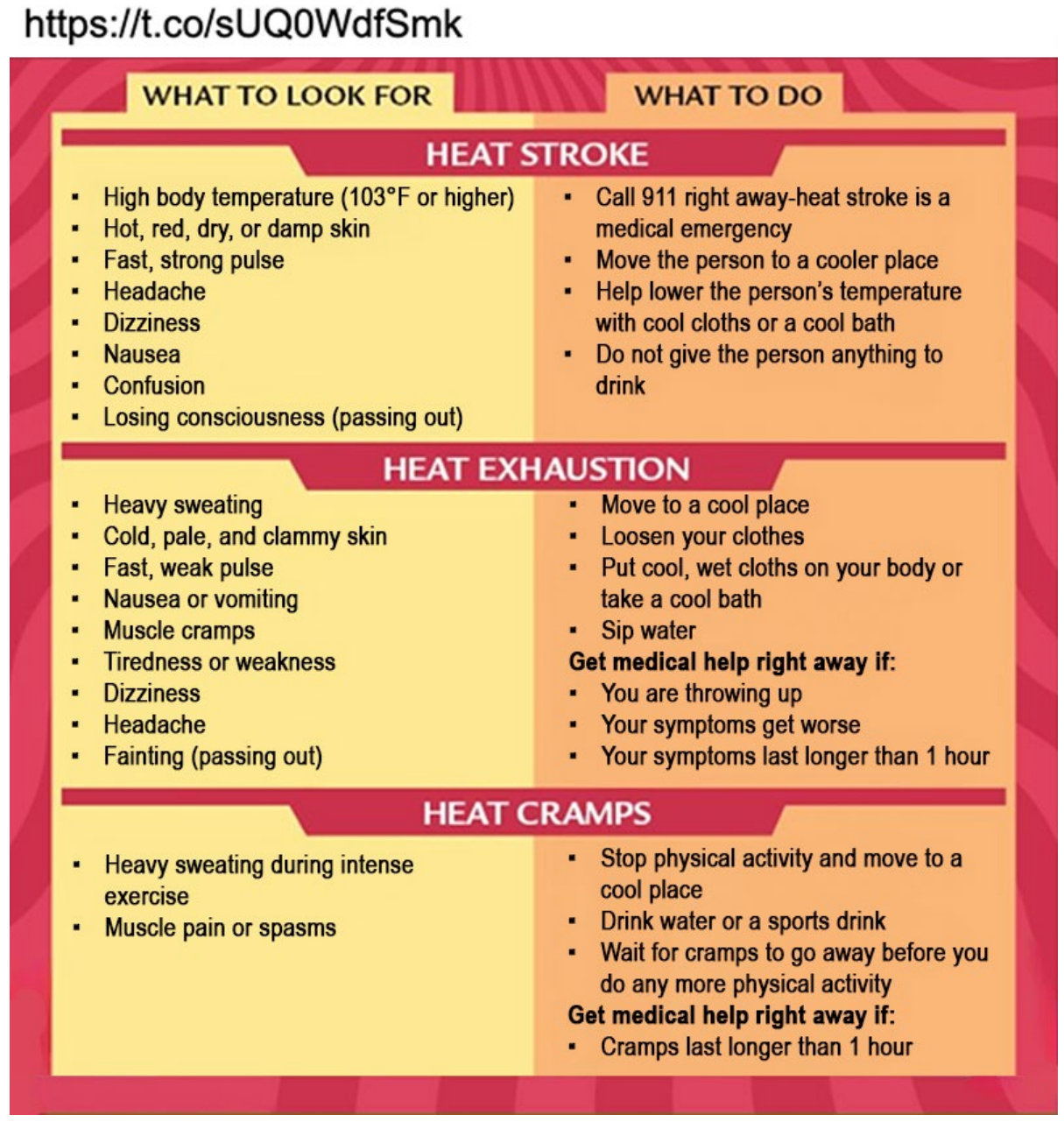

Health

Vaccination is the only long-term solution to the #COVID19India crisis. Have you taken yours yet? 2/2 #VaccinationDrive #vaccinated #CovidVaccine #Puducherry https://t.co/QasRfFRlc3

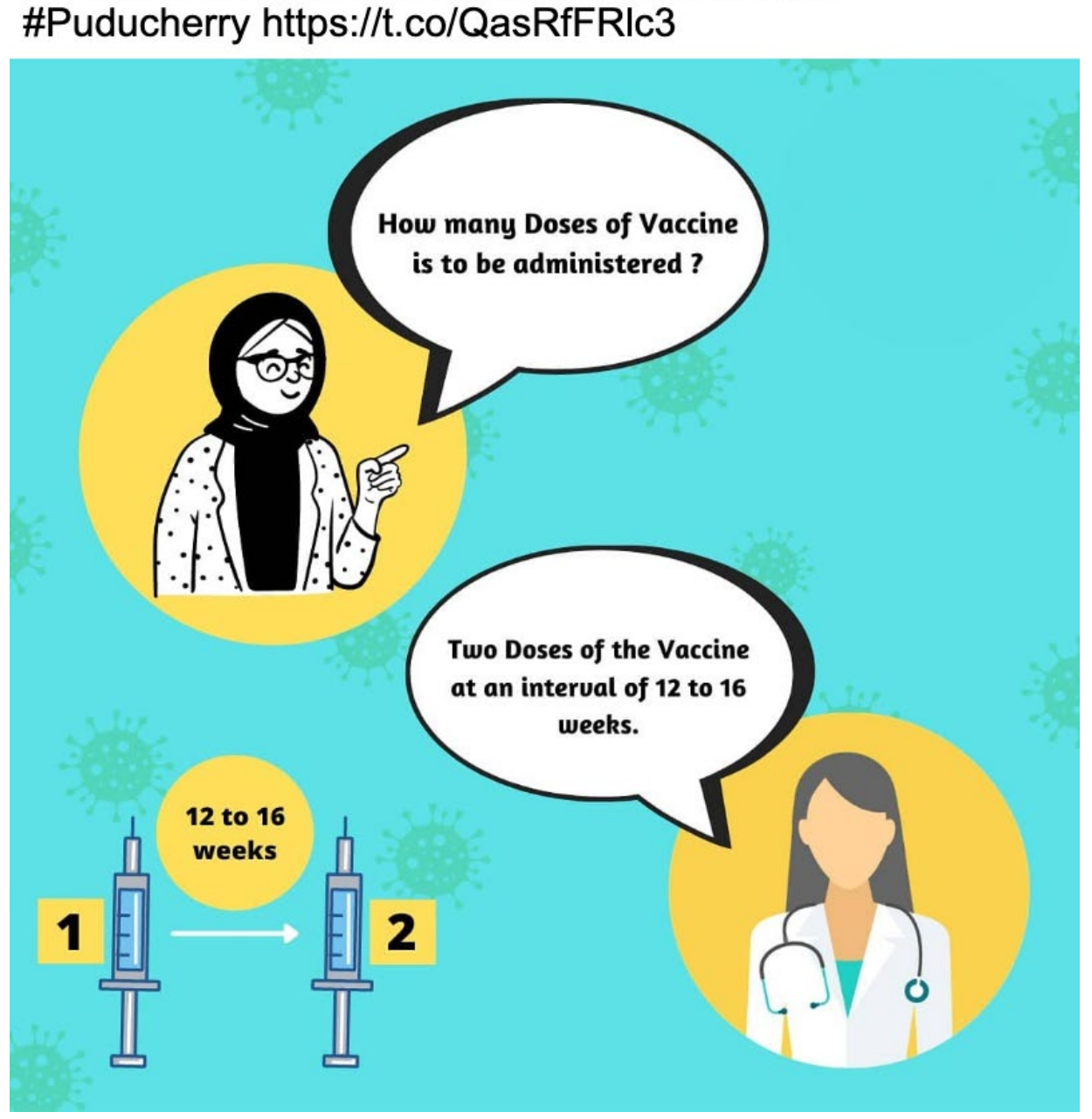

Russo-Ukrainian War

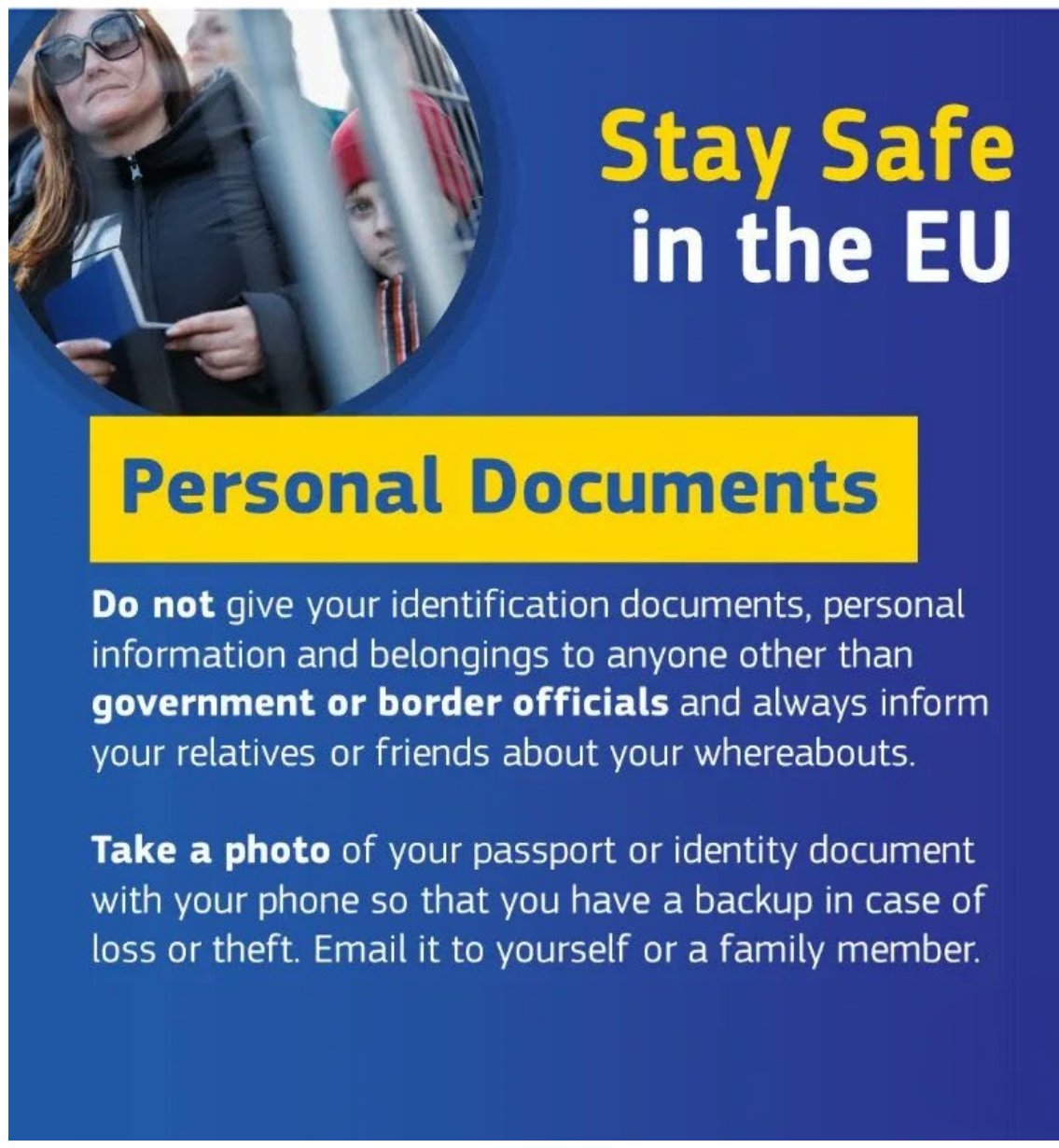

GMO

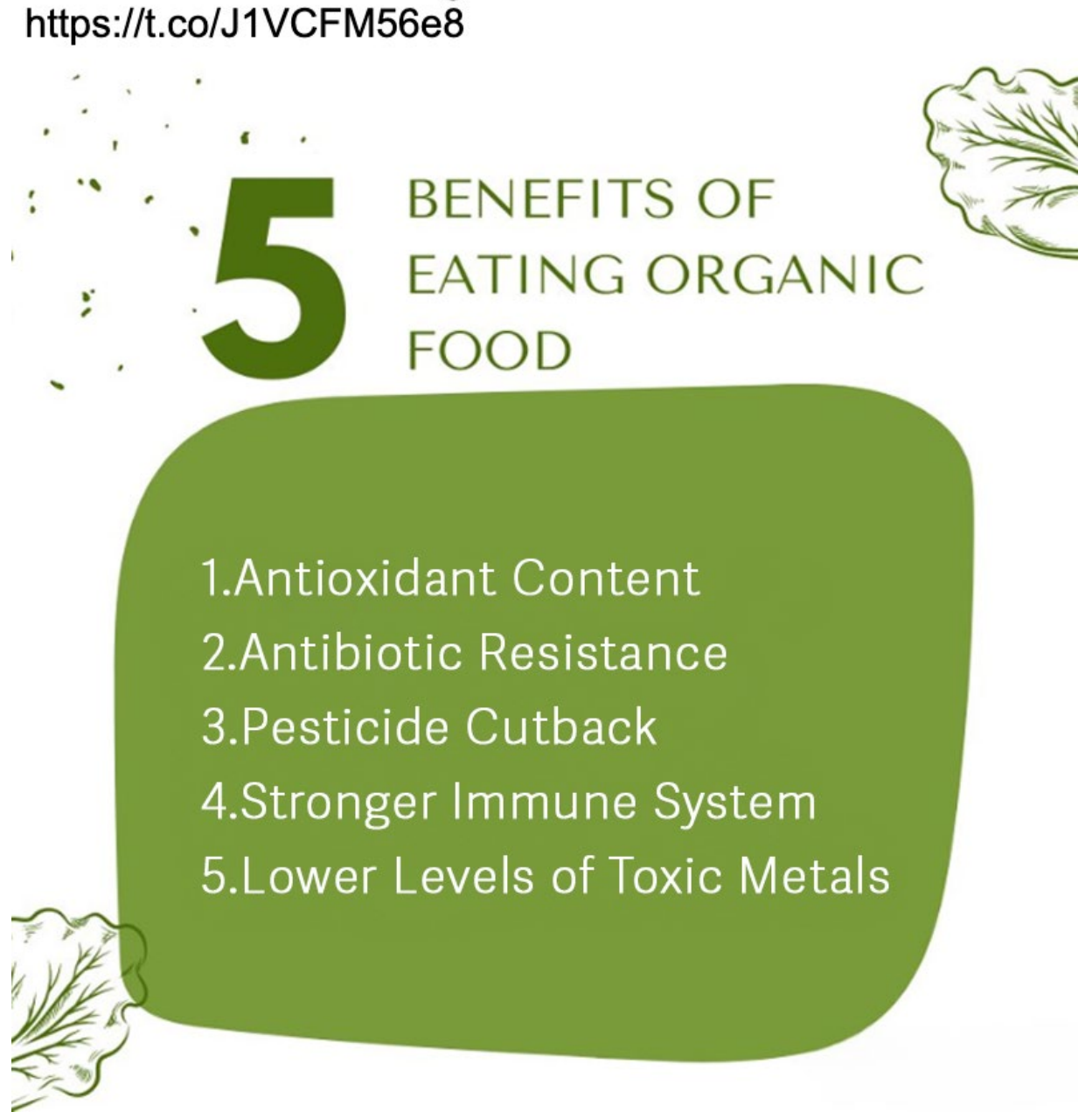

A2.2. Low Aesthetic Appeal Condition

Climate

Health

Pay attention people. Heat is the deadliest form of weather. Be safe!!! #HEATWAVE https://t.co/sUQ0WdfSmk

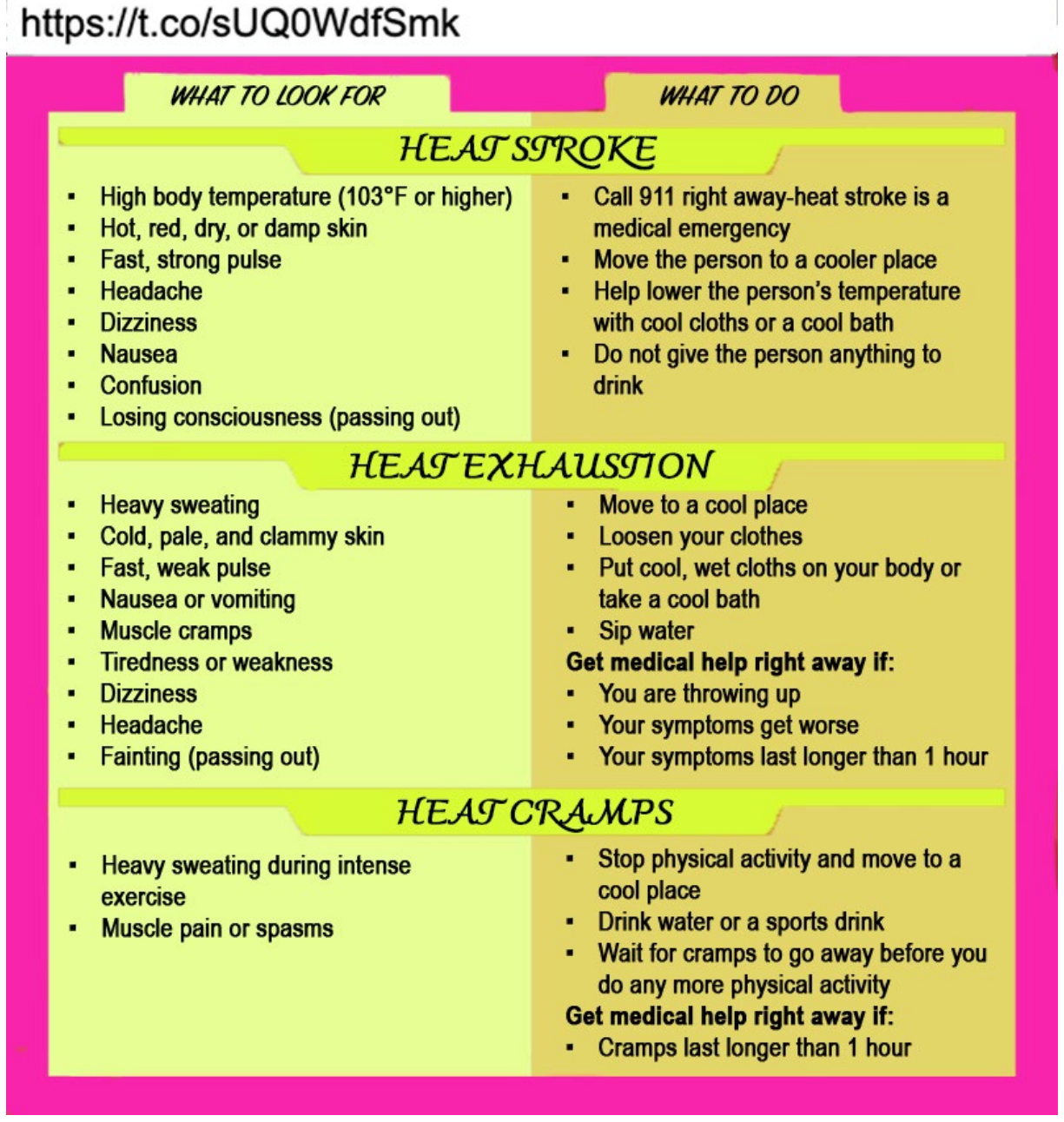

Vaccination is the only long-term solution to the #COVID19India crisis. Have you taken yours yet? 2/2 #VaccinationDrive #vaccinated #CovidVaccine #Puducherry https://t.co/QasRfFRlc3

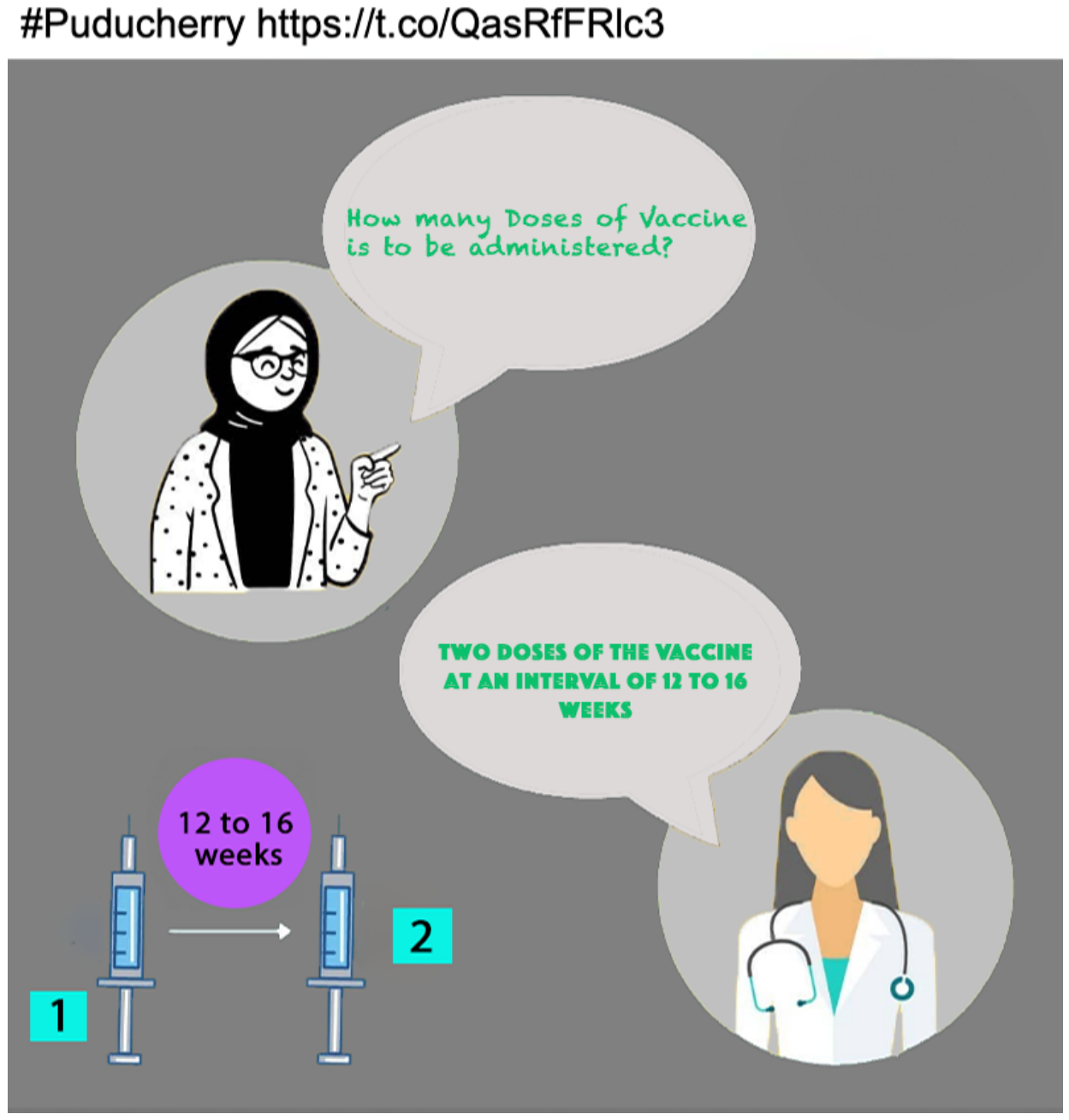

Russo-Ukrainian War

If you are fleeing Ukraine and coming to the EU, here are some useful recommendations on how to stay safe while on the move. Info on: 📄 Documents 🏢 Accommodation 🧰 Jobs 🛡 Safety from trafficking 👉 https://t.co/FYNJ5JmmIr #StandWithUkraine https://t.co/Fl3r46Bmfq

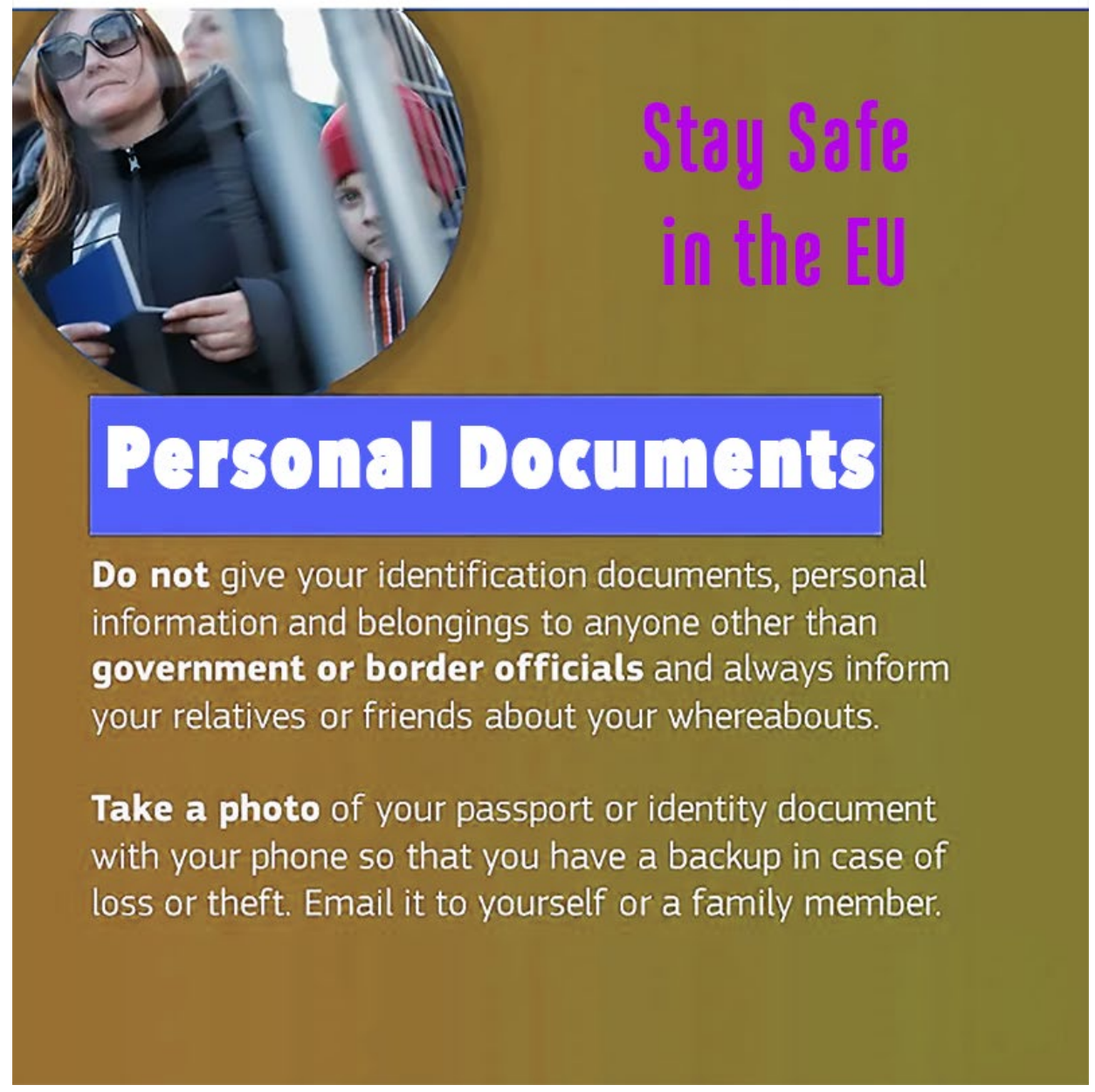

GMO

Natural food nourishes the body, not deprives it.🌿 #selfcare #wellness #organic #nutrition #NonGMO https://t.co/J1VCFM56e8

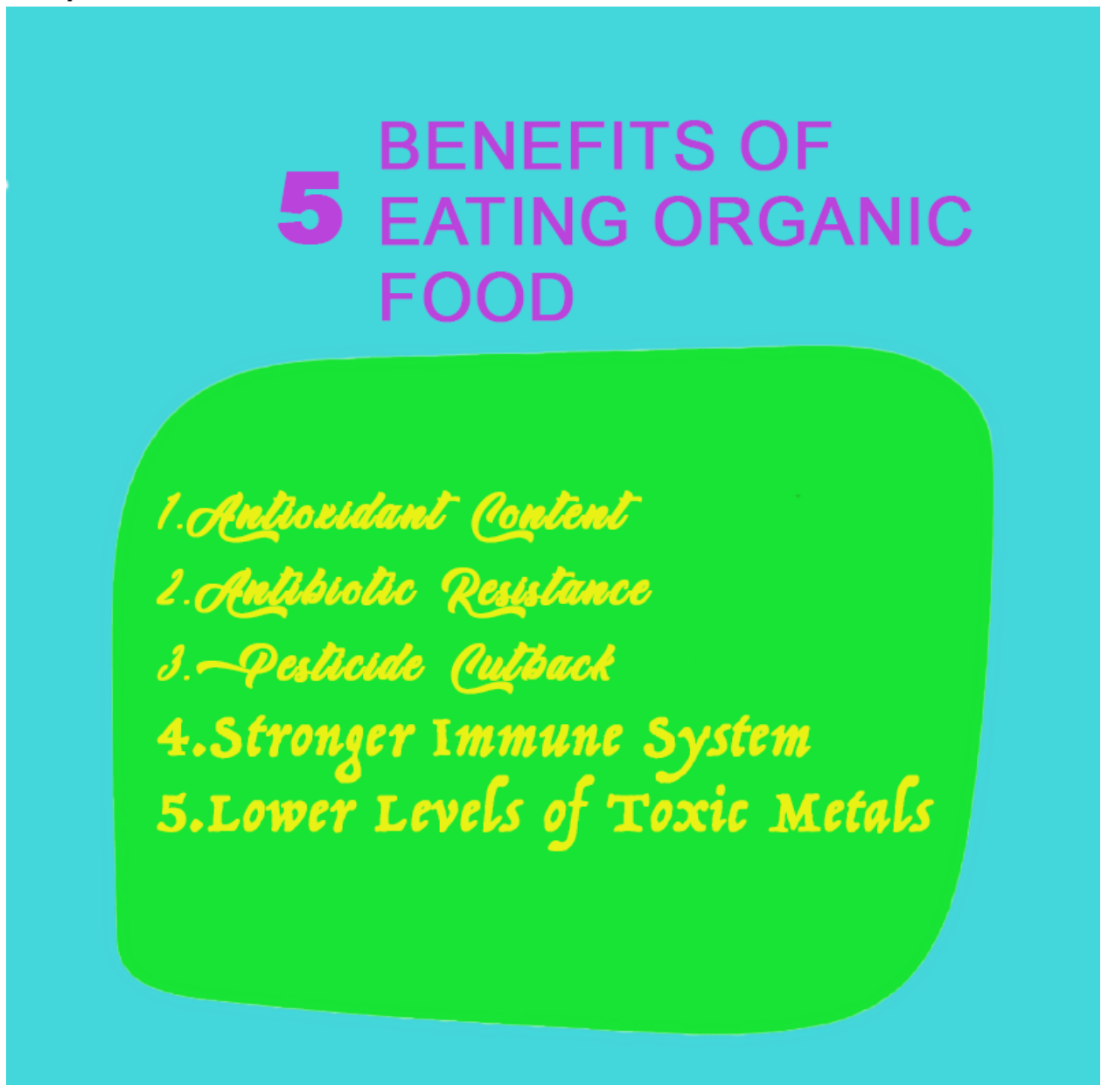

A2.3. Low Production Quality Condition

Climate

Pay attention people. Heat is the deadliest form of weather. Be safe!!! #HEATWAVE https://t.co/sUQ0WdfSmk

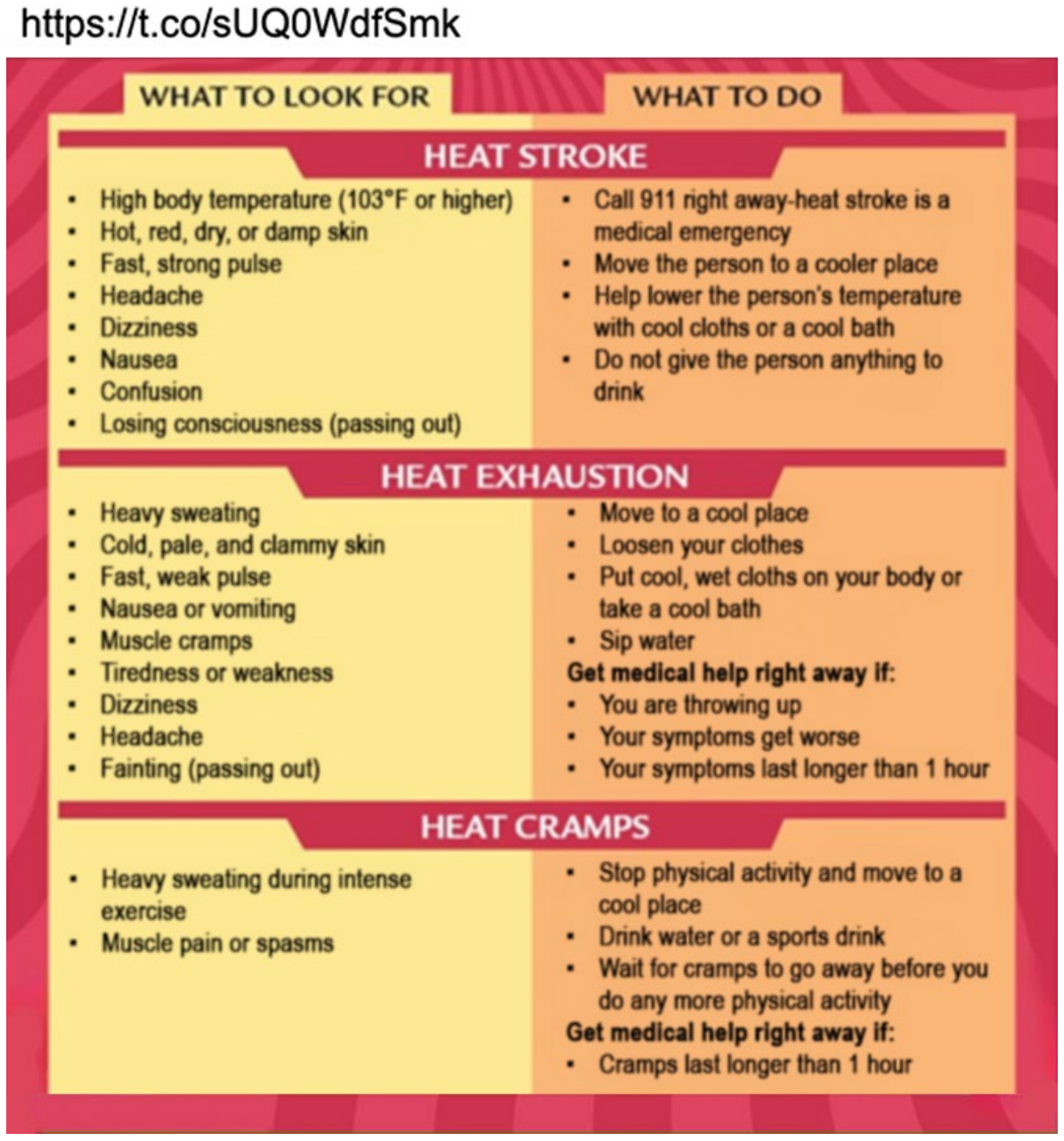

Health

Vaccination is the only long-term solution to the #COVID19India crisis. Have you taken yours yet? 2/2 #VaccinationDrive #vaccinated #CovidVaccine #Puducherry https://t.co/QasRfFRlc3

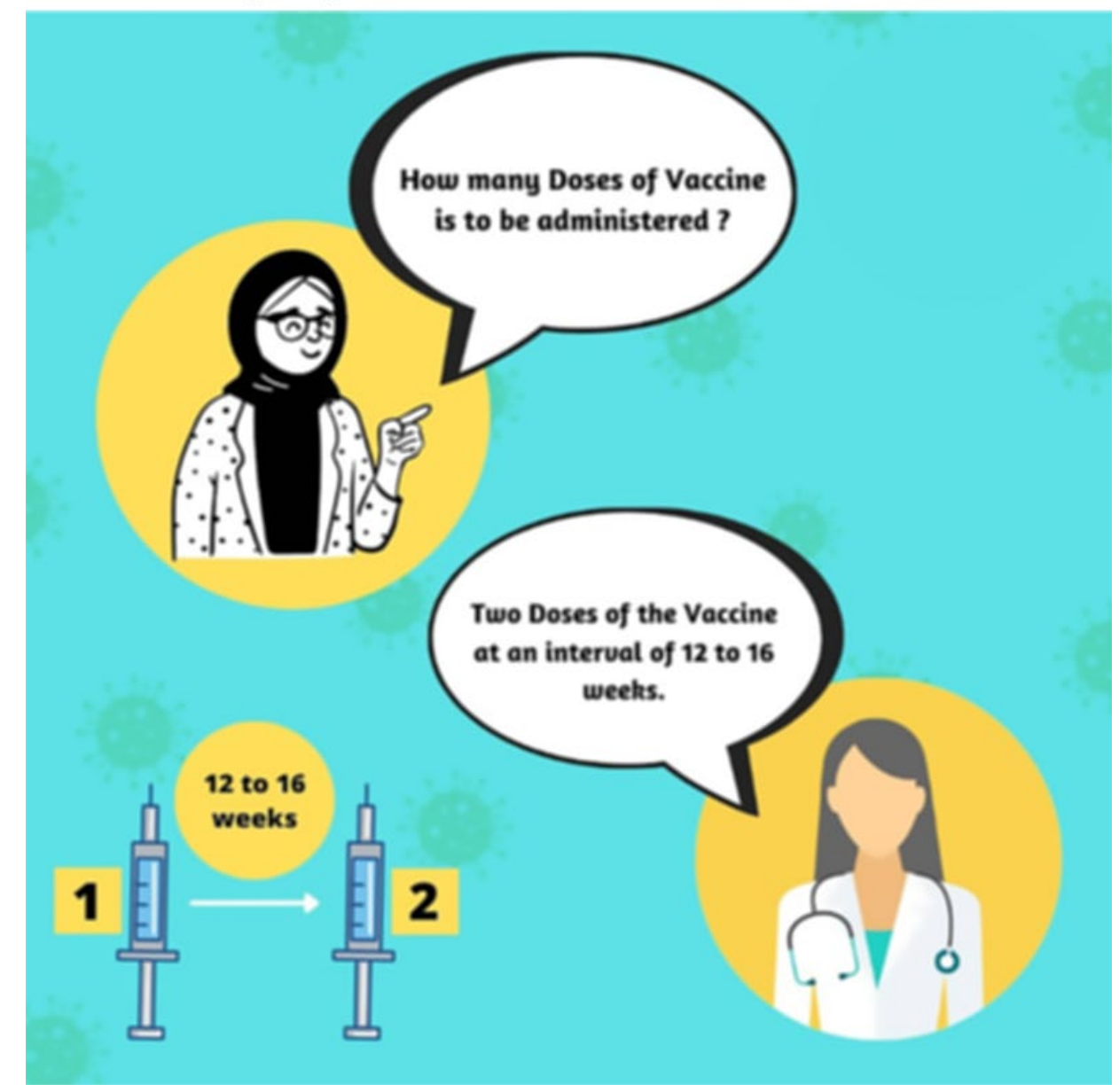

Russo-Ukrainian War

If you are fleeing Ukraine and coming to the EU, here are some useful recommendations on how to stay safe while on the move. Info on: 📄 Documents 🏢 Accommodation 🧰 Jobs 🛡 Safety from trafficking 👉 https://t.co/FYNJ5JmmIr #StandWithUkraine https://t.co/FI3r46Bmfq

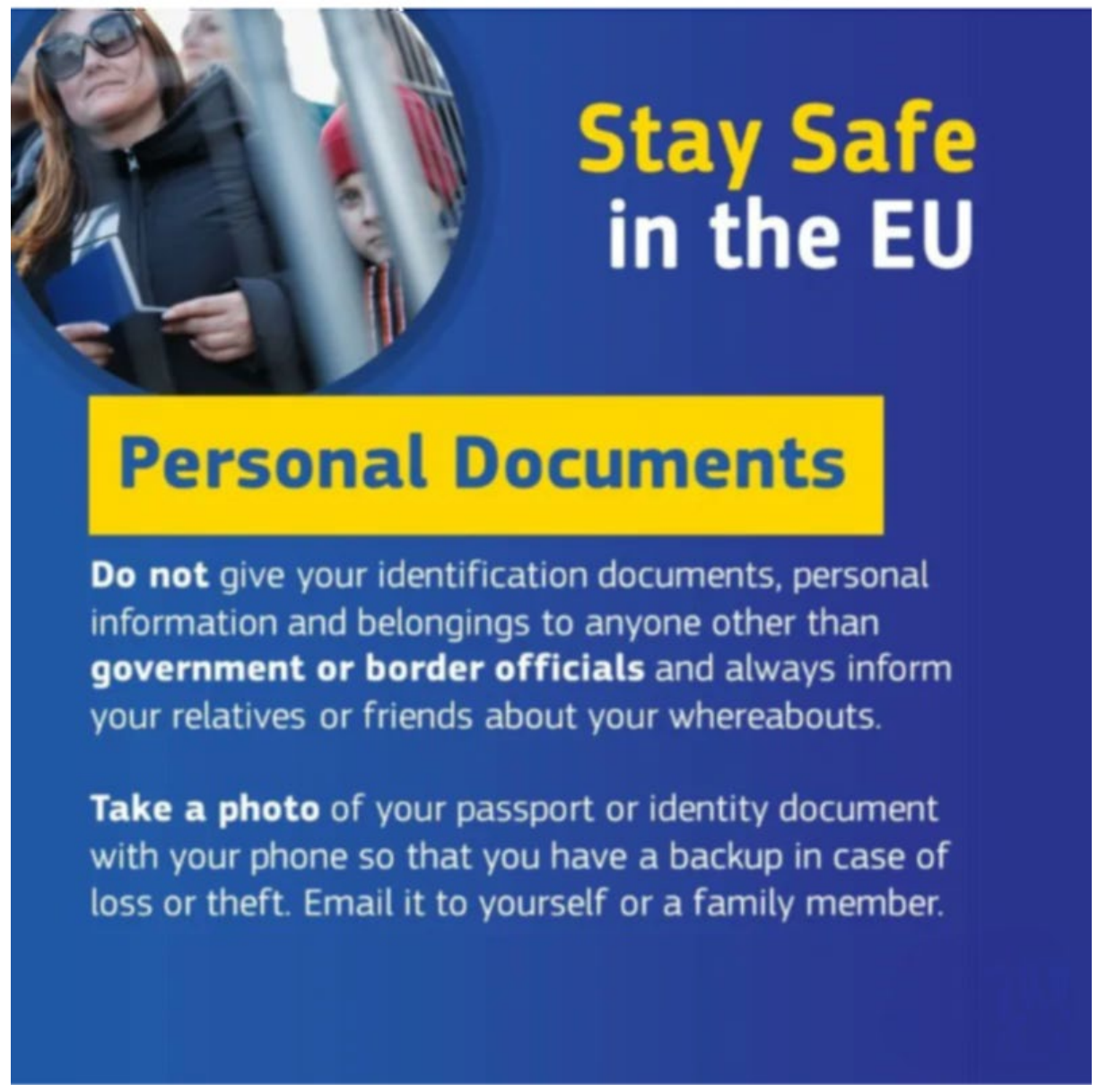

GMO

Natural food nourishes the body, not deprives it.🌿 #selfcare #wellness #organic #nutrition #NonGMO https://t.co/J1VCFM56e8

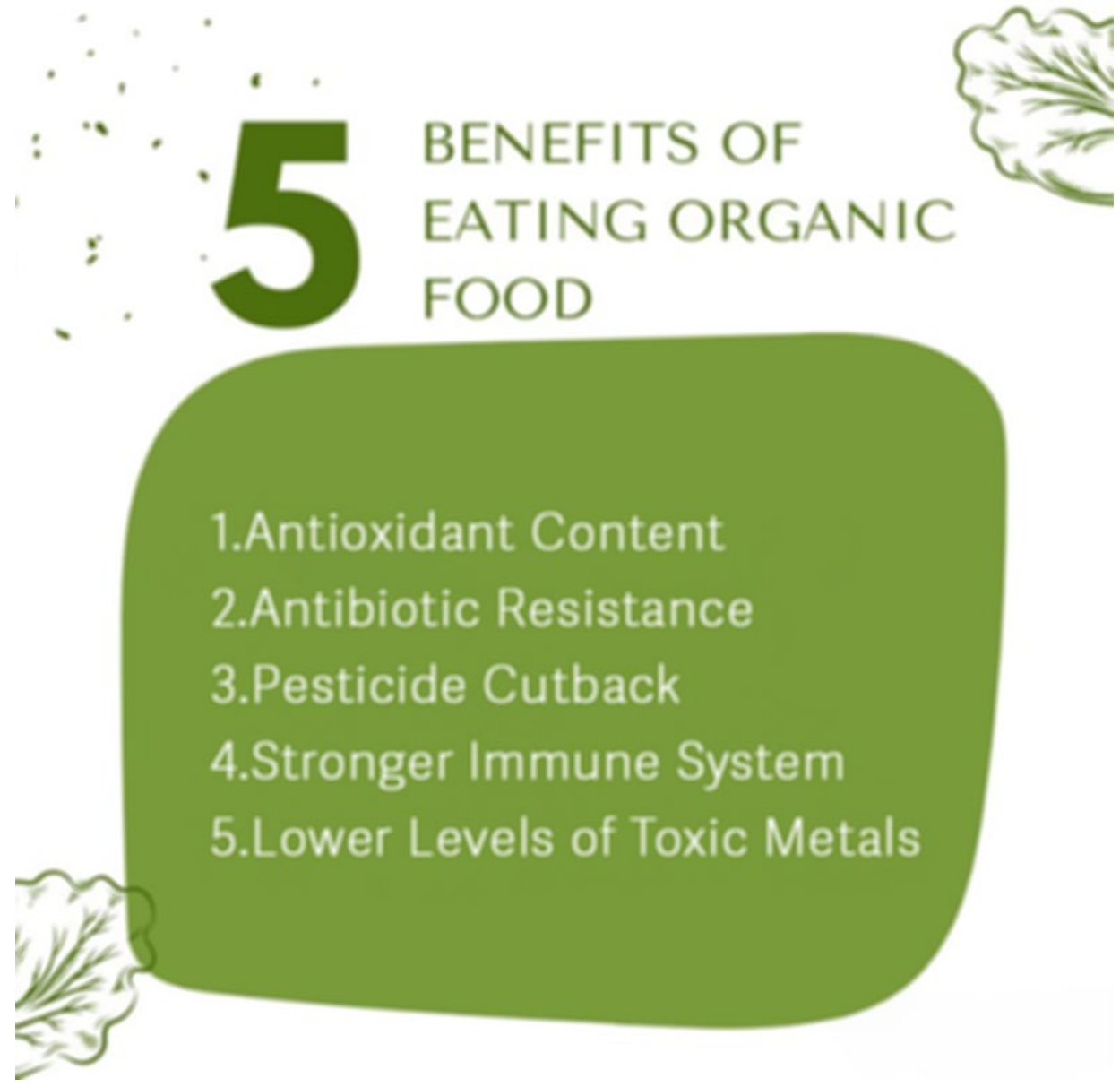

**A3. Data Visualizations**

A3.1. Original Condition

Climate

@jamieclimate For context #greennewdeal Poor #energy planning has US #economy based ~80% NONRENEWABLE #fossilfuels #renewableenergy ~12% Energy/Source US https://t.co/xeRsQQNxau Need politicians #media help public energy literacy #climatechange @nytclimate https://t.co/zdJ7OadyXQ

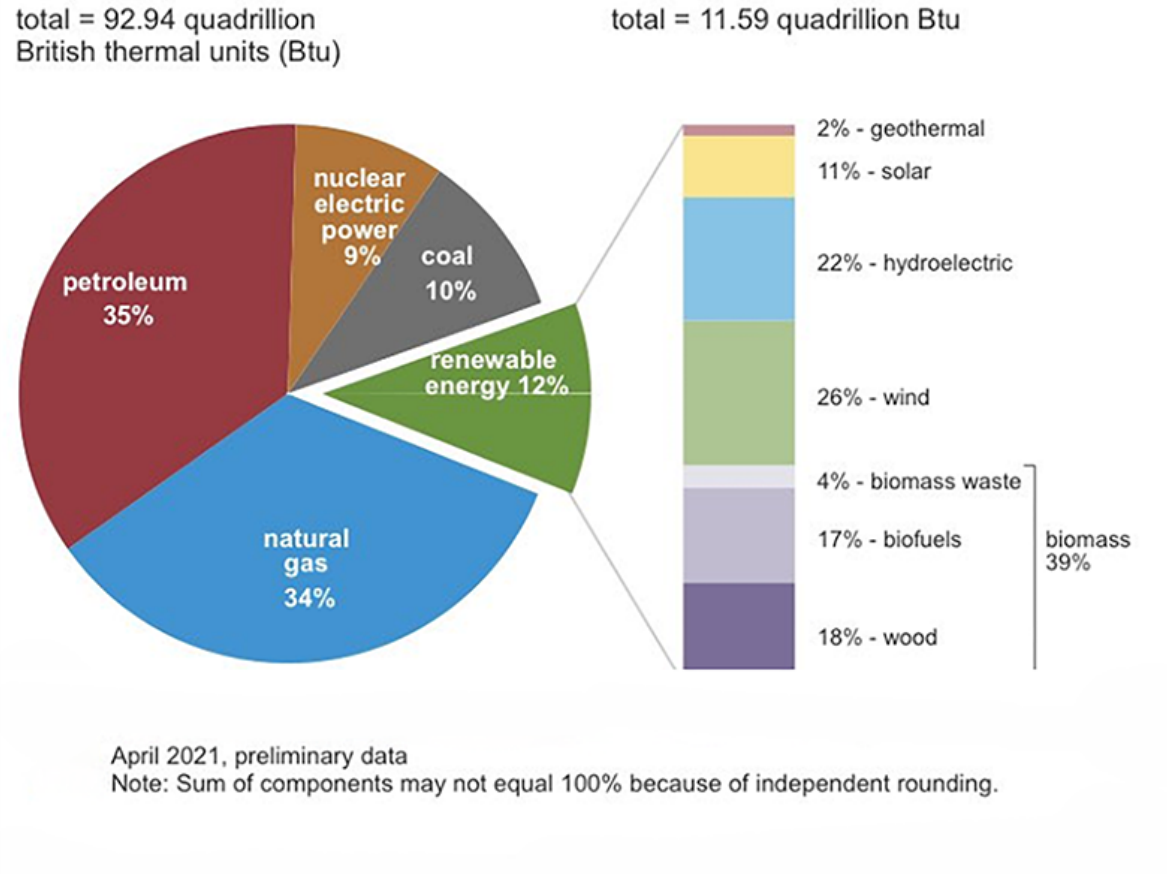

Health

When it comes to healthcare, bigger budgets don't always guarantee better coverage.#HealthSpending #PublicHealth https://t.co/QasRfFR

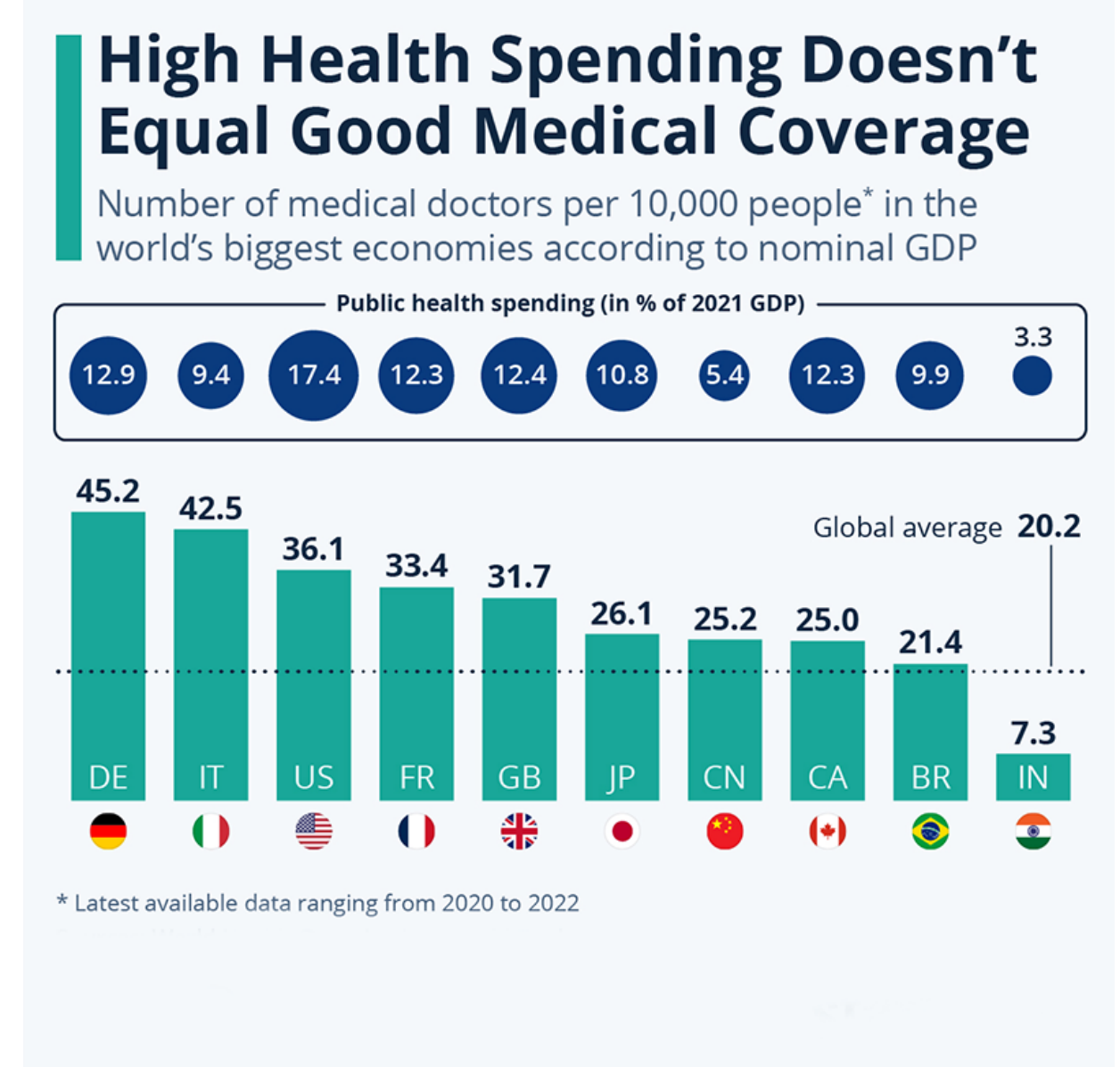

Russo-Ukrainian War

GMO

Partial numbers of the jobs lost in #russia due to the closure of western shops following the #invasion #RussianUkrainianWar #UkraineRussiaCrisis https://t.co/58meFCeNs1

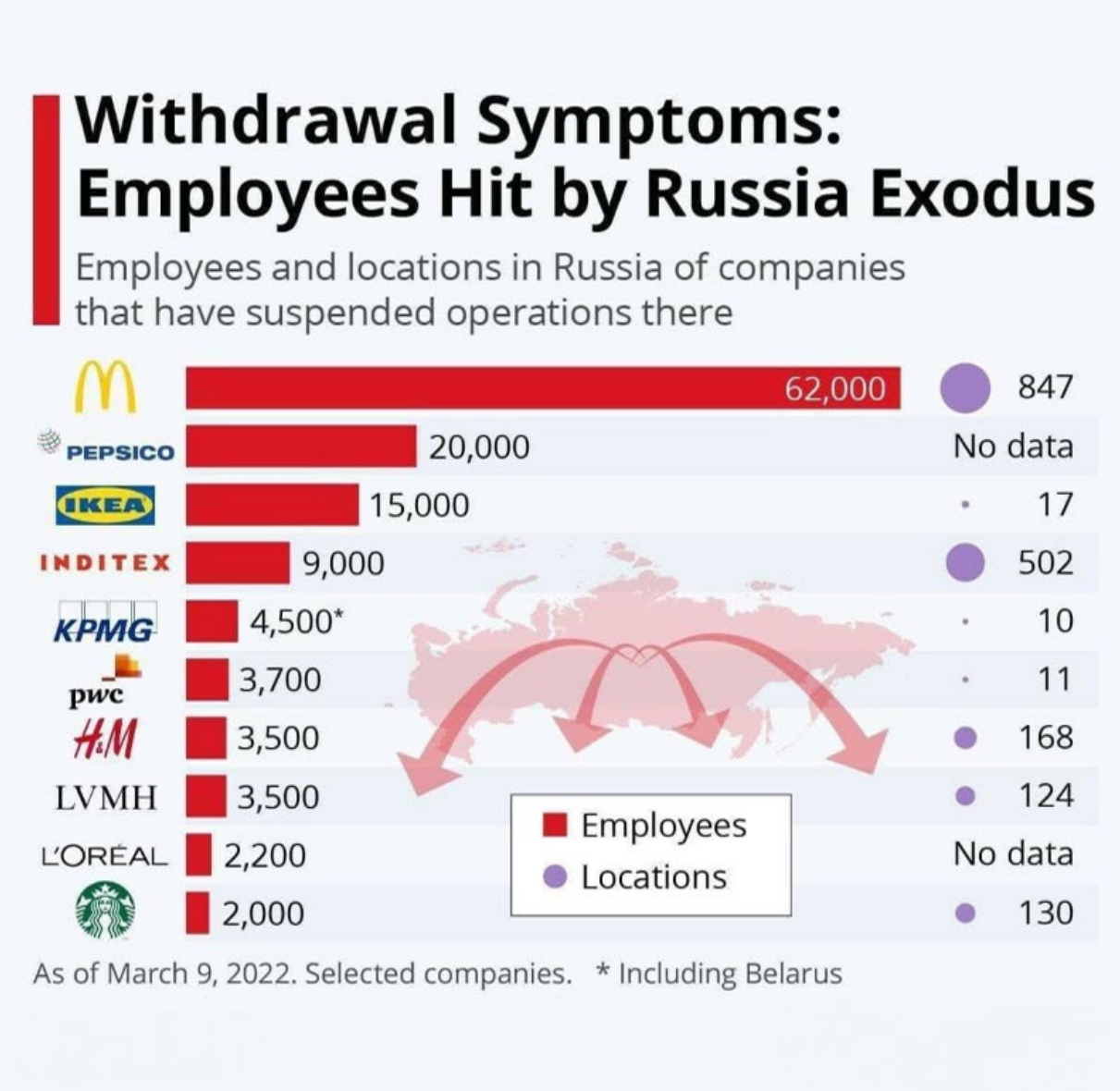

Adoption of #GeneticallyEngineered Crops in the U.S. Read more: http://shorturl.at/jkvBP@APCoAB @agbiotech @AgBioWorld @GMOinfoEU @isaaa_org @ICGEB @ISAAASciSpeaks @gmopundit

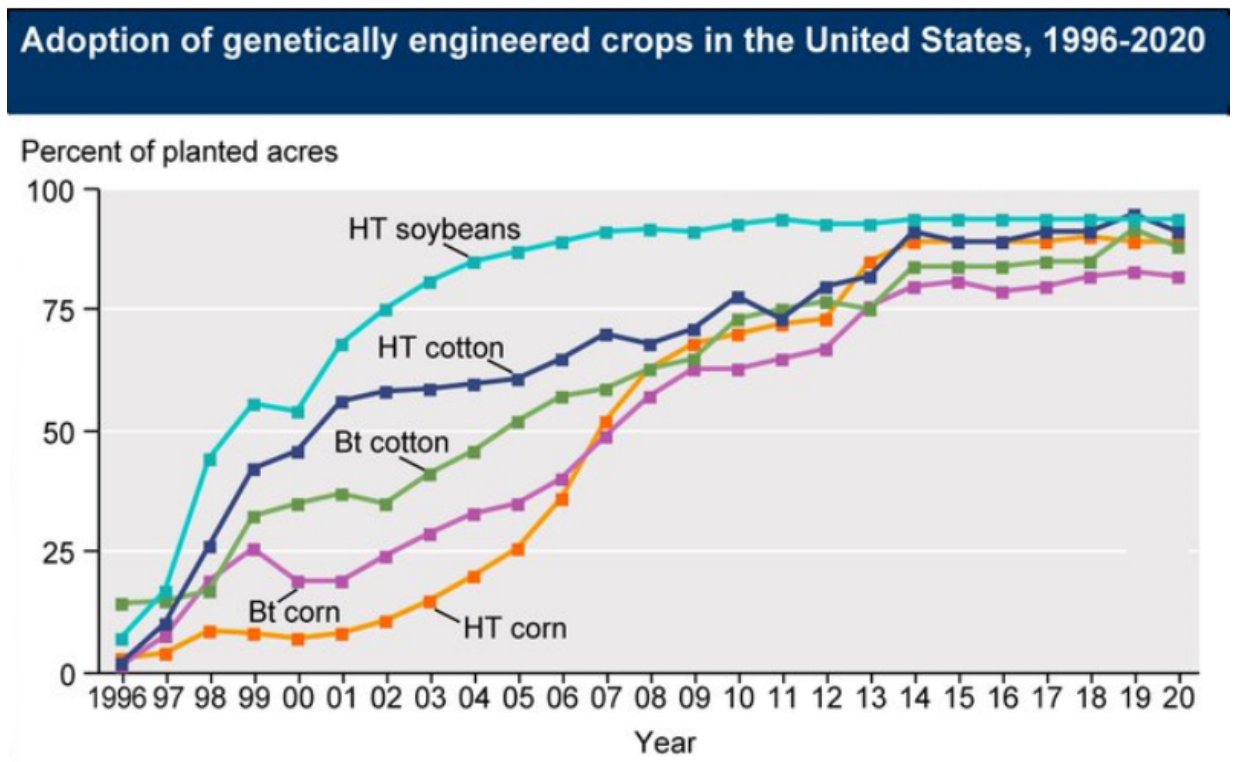

A3.2. Low Aesthetic Appeal Condition

Climate

Health

@jamieclimate For context #greennewdeal Poor #energy planning has US #economy based ~80% NONRENEWABLE #fossilfuels #renewableenergy ~12% Energy/Source US https://t.co/xeRsQQNxau Need politicians #media help public energy literacy #climatechange @nytclimate https://t.co/zdJ7OadyXQ

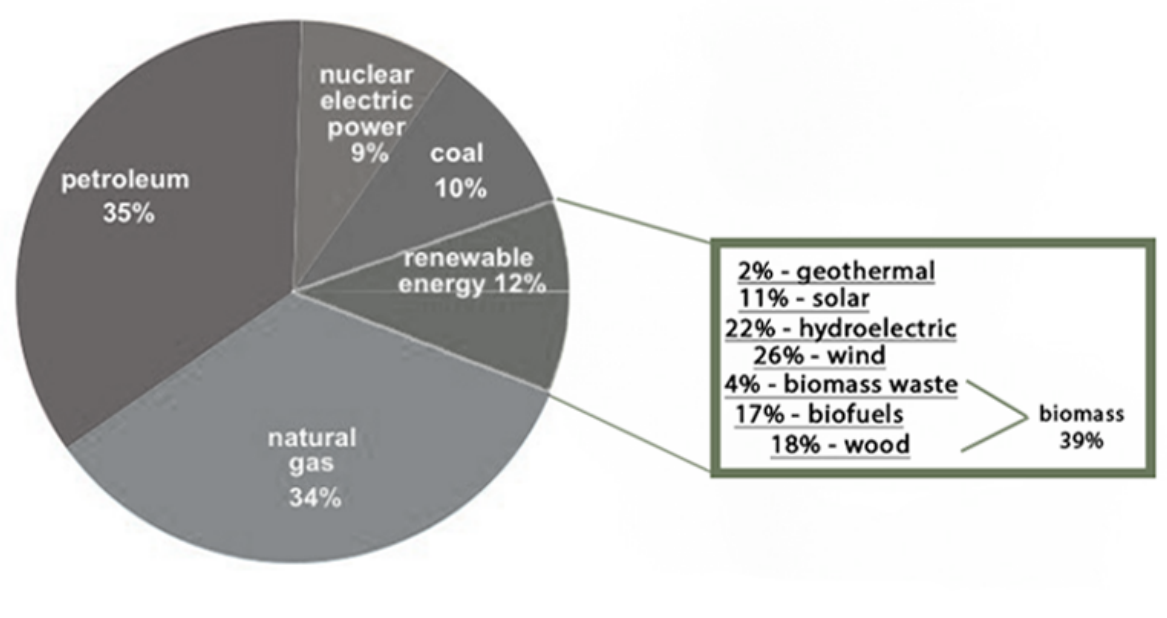

When it comes to healthcare, bigger budgets don't always guarantee better coverage.#HealthSpending #PublicHealth https://t.co/QasRfFR

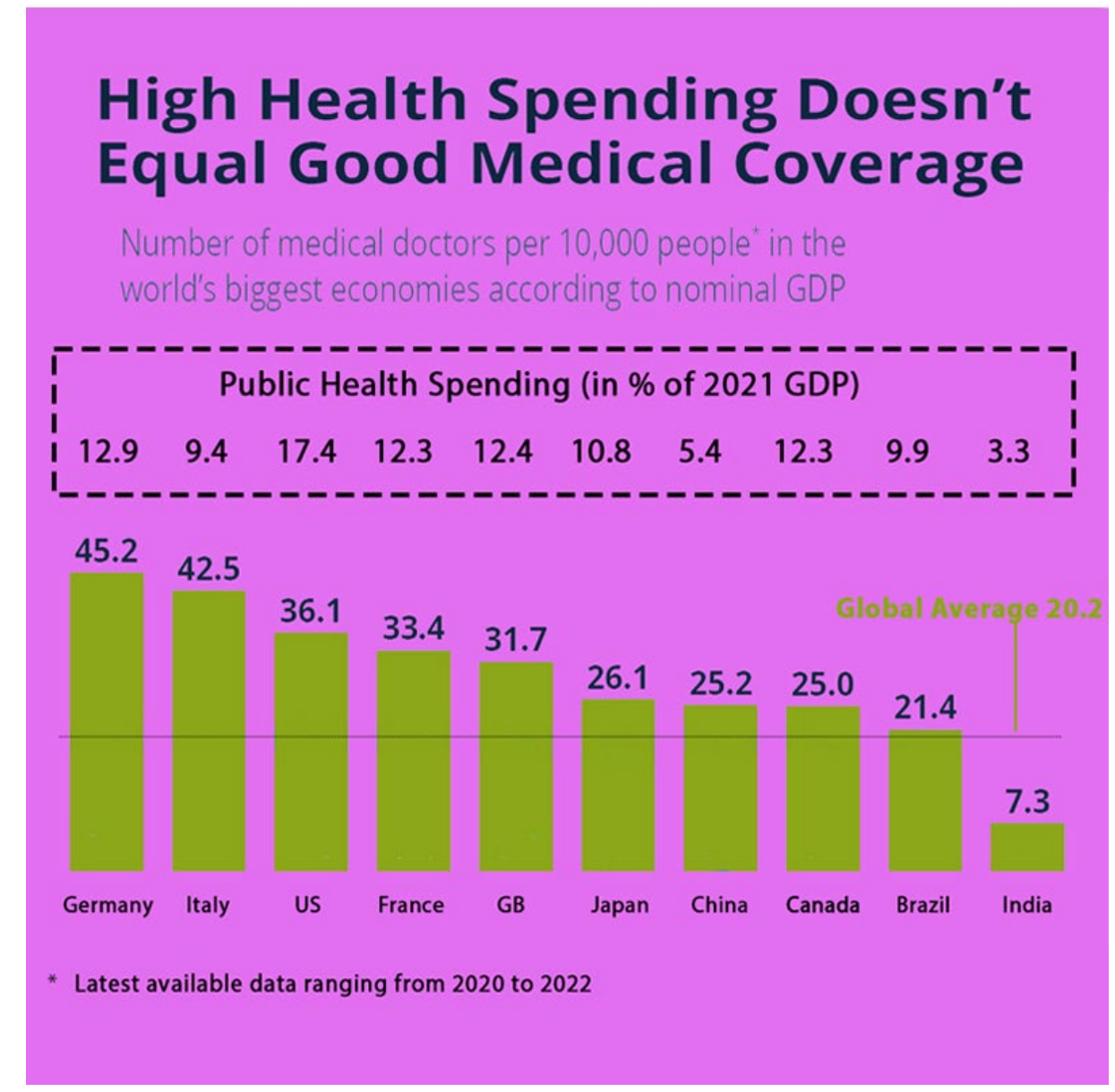

Russo-Ukrainian War

Partial numbers of the jobs lost in #russia due to the closure of western shops following the #invasion #RussianUkrainianWar #UkraineRussiaCrisis https://t.co/58meFCeNs1

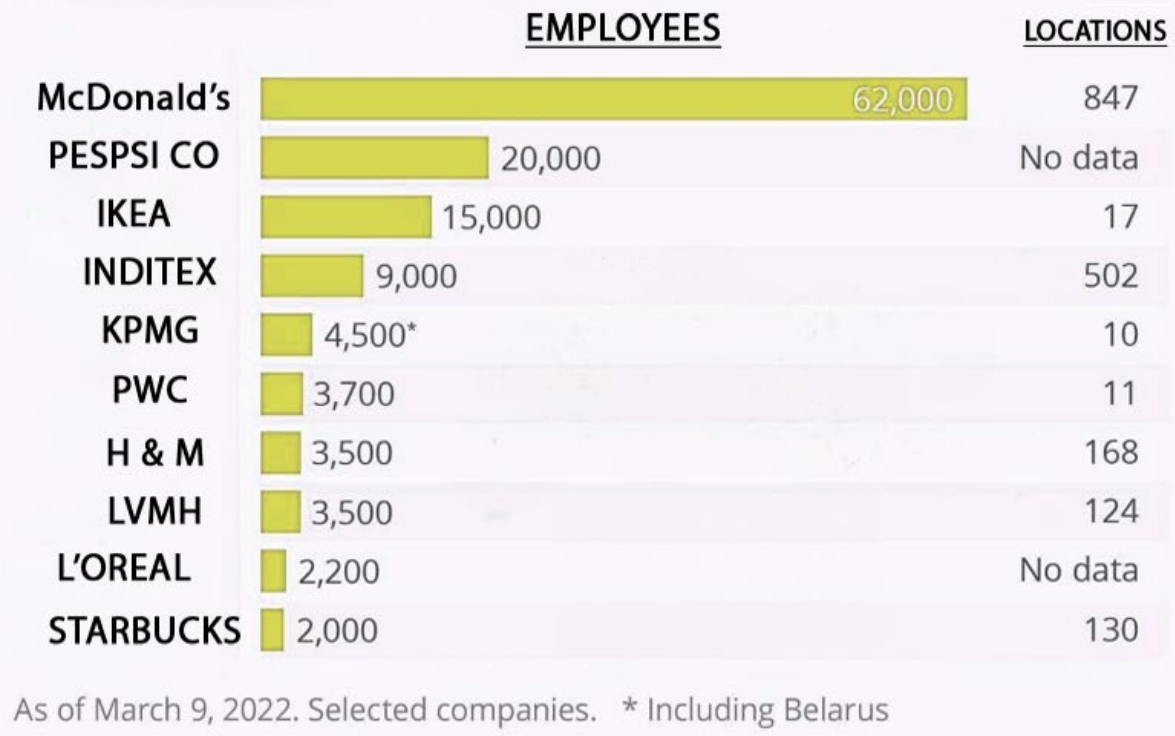

GMO

Adoption of #GeneticallyEngineered Crops in the U.S. Read more: http://shorturl.at/jkvBP@APCoAB @agbiotech @AgBioWorld @GMOinfoEU @isaaa_org @ICGEB @ISAAASciSpeaks @gmopundit

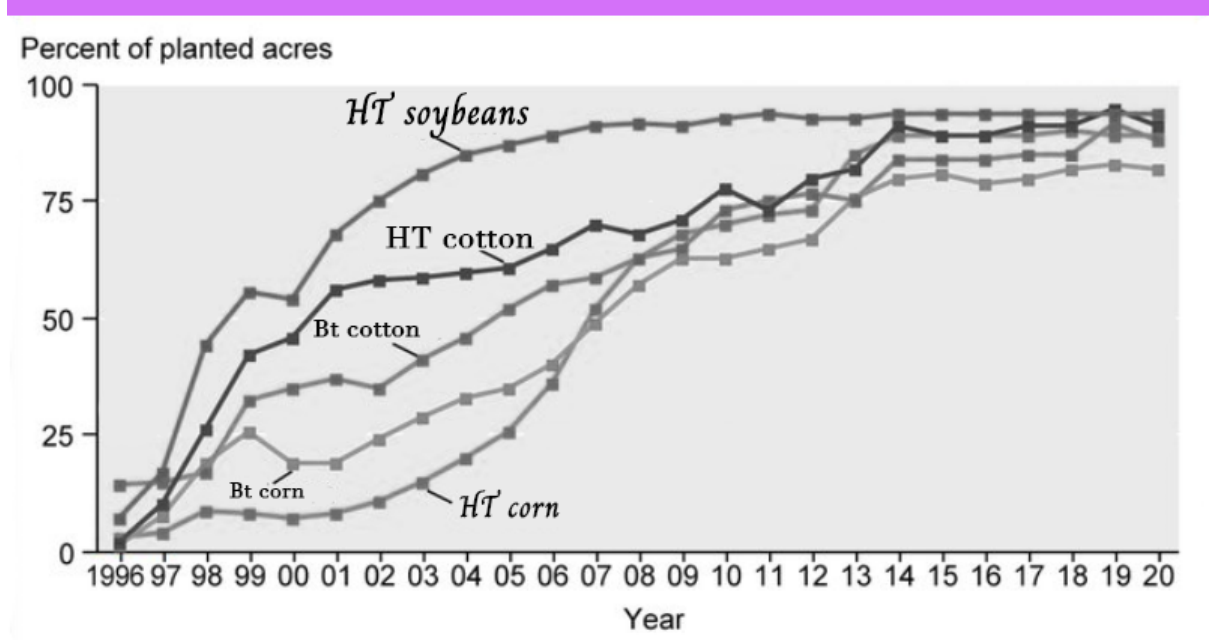

A3.3. Low Production Quality Condition

Climate

@jamieclimate For context #greennewdeal Poor #energy planning has US #economy based ~80% NONRENEWABLE #fossilfuels #renewableenergy ~12% Energy/Source US https://t.co/xeRsQQNxau Need politicians #media help public energy literacy #climatechange @nytclimate https://t.co/zdJ7OadyXQ

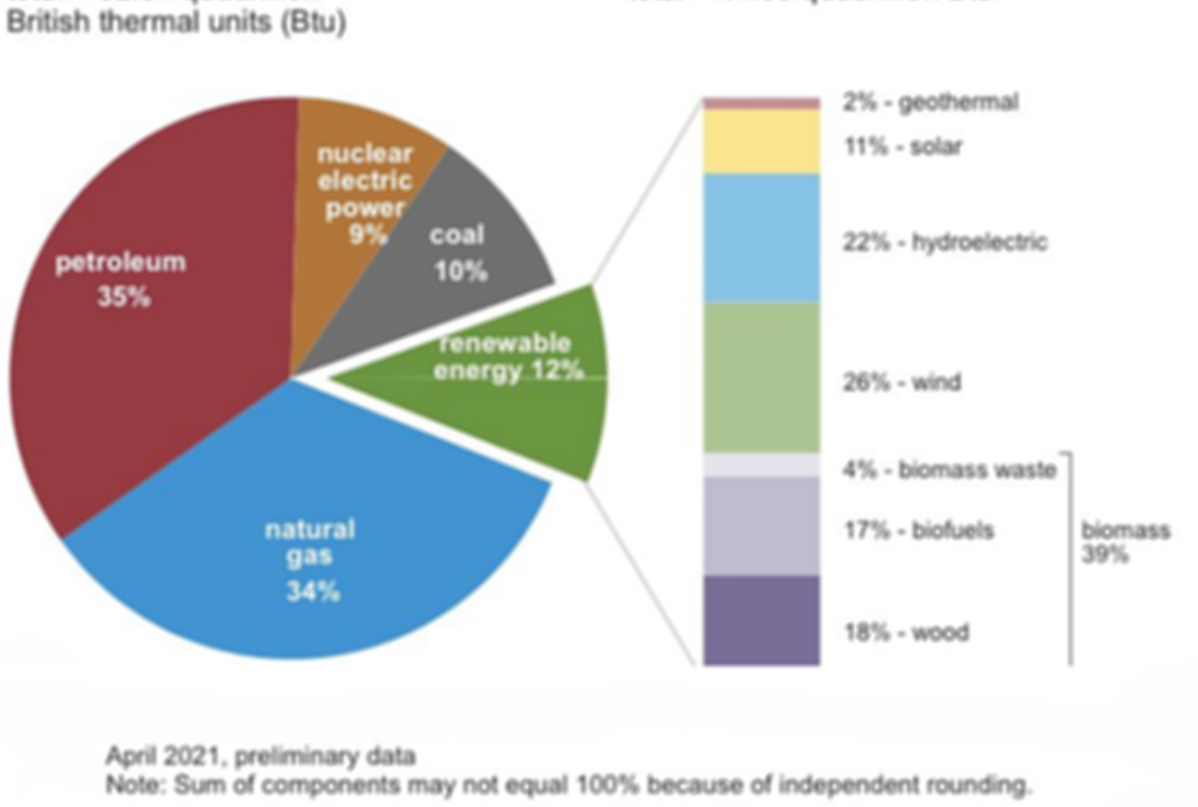

Health

When it comes to healthcare, bigger budgets don't always guarantee better coverage.#HealthSpending #PublicHealth https://t.co/QasRfFR

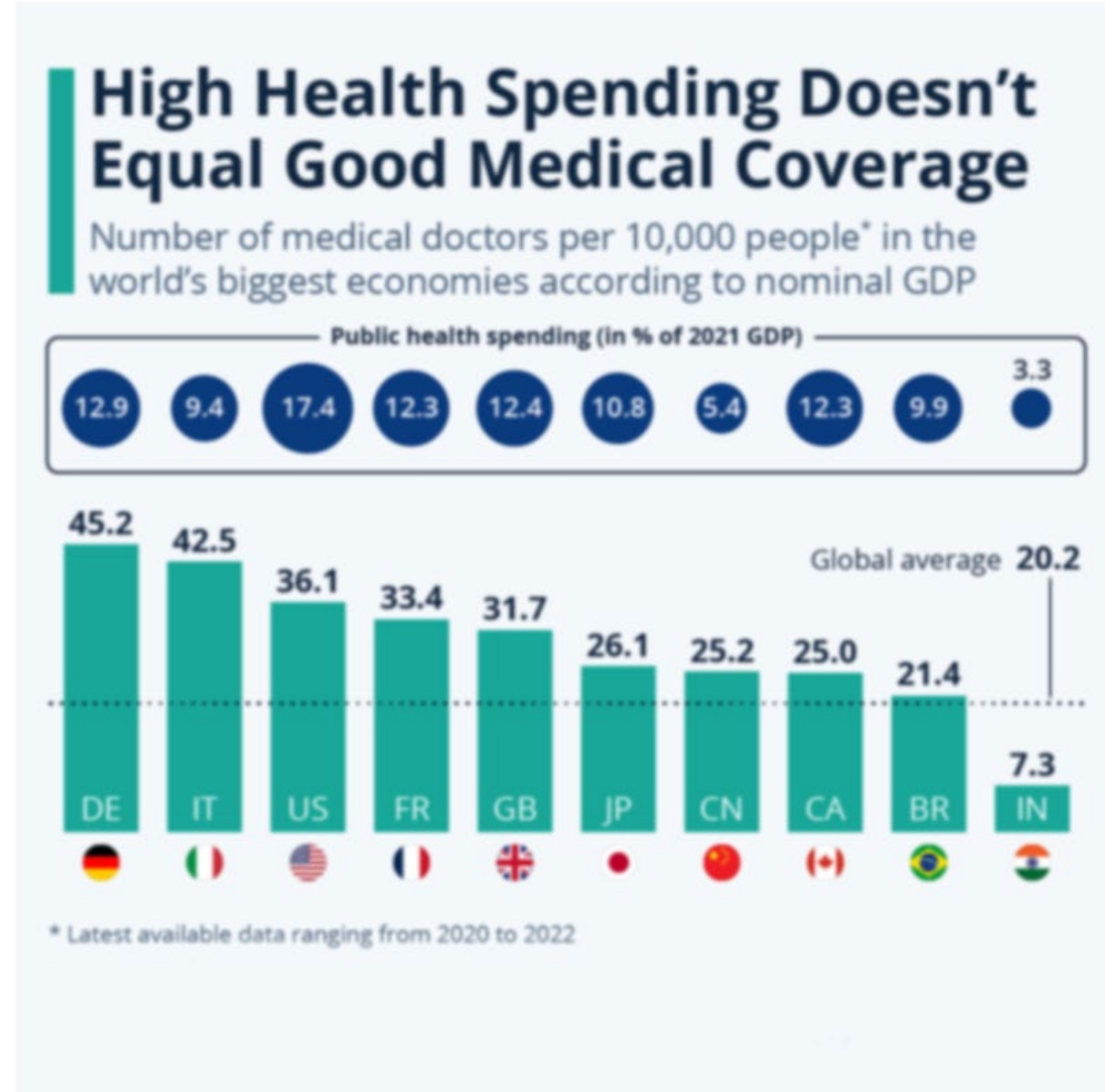

Russo-Ukrainian War

Partial numbers of the jobs lost in #russia due to the closure of western shops following the #invasion #RussianUkrainianWar #UkraineRussiaCrisis https://t.co/58meFCeNs1

**Withdrawal Symptoms: Employees Hit by Russia Exodus**

Employees and locations in Russia of companies that have suspended operations there

|  | Employees | Locations |
|---|---|---|
| McDonald's | 62,000 | 847 |
| PEPSICO | 20,000 | No data |
| IKEA | 15,000 | 17 |
| INDITEX | 9,000 | 502 |
| KPMG | 4,500* | 10 |
| pwc | 3,700 | 11 |
| H&M | 3,500 | 168 |
| LVMH | 3,500 | 124 |
| L'OREAL | 2,200 | No data |
| Starbucks | 2,000 | 130 |

As of March 9, 2022. Selected companies. * Including Belarus

GMO

Adoption of #GeneticallyEngineered Crops in the U.S. Read more: http://shorturl.at/jkvBP@APCoAB @agbiotech @AgBioWorld @GMOinfoEU @isaaa_org @ICGEB @ISAAASciSpeaks @gmopundit

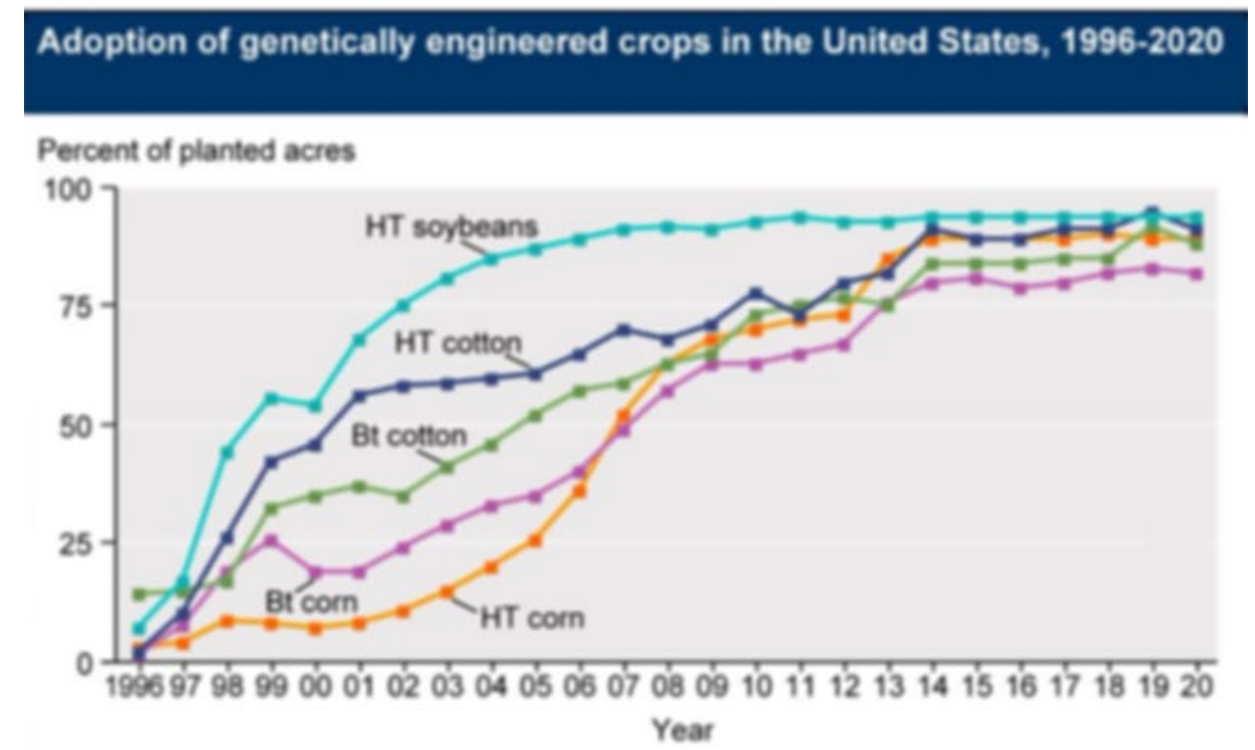

**A4. Text-only Posts**

**Climate**

Around the world, poor and vulnerable communities are facing multiple crises at once. Sadly, they have contributed least to the #climatecrisis but, are being hit the hardest. https://t.co/9mxf4t6J5T

Pay attention people. Heat is the deadliest form of weather. Be safe!!! #HEATWAVE https://t.co/sUQ0WdfSmk

@jamieclimate For context #greennewdeal Poor #energy planning has US #economy based ~80% NONRENEWABLE #fossilfuels #renewableenergy ~12% Energy/Source US https://t.co/xeRsQQNxau Need politicians #media help public energy literacy #climatechange @nytclimate https://t.co/zdJ7OadyXQ

**Health**

Vaccines💉 are the world's safest way to protect children from serious diseases such as polio, measles and smallpox. They help children grow up healthy and happy. #VaccinesWork #ForEveryChild @gavi @Sida @WHO @SanteTchad @OnuTchad https://t.co/Km3iBmkEWc

Vaccination is the only long-term solution to the #COVID19India crisis. Have you taken yours yet? 2/2 #VaccinationDrive #vaccinated #CovidVaccine #Puducherry https://t.co/QasRfFRlc3

When it comes to healthcare, bigger budgets don't always guarantee better coverage.#HealthSpending #PublicHealth https://t.co/QasRfFR

**Russo-Ukrainian War**

Russia said its #missile cruiser #Moskva has been seriously damaged after an ammunition explosion, Interfax has reported. The #crew of the Moskva has been evacuated and the cause of the fire that prompted the explosion is being investigated. 🇷🇺🇺🇦 #RussianUkrainianWar https://t.co/cFOCIHK4tQ

| If you are fleeing Ukraine and coming to the EU, here are some useful recommendations on how to stay safe while on the move. Info on: 📄 Documents 🏢 Accommodation 🧰 Jobs 🛡️ Safety from trafficking 👉 https://t.co/FYNJ5JmmIr #StandWithUkraine https://t.co/FI3r46Bmfq |
|---|
| Partial numbers of the jobs lost in #russia due to the closure of western shops following the #invasion #RussianUkrainianWar #UkraineRussiaCrisis https://t.co/58meFCeNs1 |

**GMOs**

| Just how bad are GMO's? Read this article for what experts say about GMO's and if they're safe to consume. https://t.co/WAgyBsMCl7 #gmo #nutrition #health https://t.co/DilcKUJBCF |
|---|
| Natural food nourishes the body, not deprives it.🌿 #selfcare #wellness #organic #nutrition #NonGMO https://t.co/J1VCFM56e8 |
| Adoption of #GeneticallyEngineered Crops in the U.S. Read more: http://shorturl.at/jkvBP@APCoAB @agbiotech @AgBioWorld @GMOinfoEU @isaaa_org @ICGEB @ISAAASciSpeaks @gmopundit |

**B. Pilot Tests**

Potential stimuli were first chosen from a repository of social media posts rated by crowdsourced workers, followed by multiple rounds of pilot tests and modifications to create the final stimuli to be used in the main experiment. The first pilot test was done using a Large Language Model (LLM), GPT-4o, to test the effectiveness of manipulations in aesthetic appeal and production quality. After the stimuli were further refined, the subsequent rounds of pilot tests were conducted with US undergraduate students as human raters before selecting the final set of stimuli for the main experiment.

**B1. Stimuli Selection**

We chose stimuli from a previous study of visual social media posts (Peng et al., 2025). Stimuli rated as medium to high in terms of aesthetic appeal and credibility on a 7-point scale were selected (see Table B1). Although the stimuli were not rated in terms of production quality, all potential stimuli had a resolution of 800 x 800 pixels at least.

**Table B1.** Crowdsourced Ratings of Shortlisted Stimuli

| | Issue Context | Format | Credibility | Aesthetic Appeal |
|---|---|---|---|---|
| 1. | Climate | Photo | 5.70 | 4.88 |
| 2. | Climate | Infographic | 5.96 | 4.33 |
| 3. | Climate | Data Visualization | 5.26 | 5.50 |
| 4. | Health | Photo | 5.90 | 5.45 |
| 5. | Health | Infographic | 5.93 | 4.74 |
| 6. | Health | Data Visualization[a] | | |
| 7. | RUWar | Photo | 4.66 | 5.33 |
| 8. | RUWar | Infographic | 5.17 | 5.07 |
| 9. | RUWar | Data Visualization | 5.15 | 5.19 |
| 10. | GMO | Photo | 4.05 | 4.50 |
| 11. | GMO | Infographic | 5.30 | 5.20 |
| 12. | GMO | Visualization | 5.20 | 3.80 |

*Notes.* The table represents the mean credibility and aesthetic appeal scores of shortlisted stimuli rated by crowdsourced workers in Peng et al. (2025), to be used in the original condition of the main experiment. Ratings were on a scale of 1 to 7 (lowest to highest). [a] No feasible data visualization for health could be found from the repository. Hence, the data visualization was selected from *statista.com* (#6).

## B2. Pilot Test Using LLM

Based on the selected stimuli for the original condition, we created the corresponding stimuli for the low aesthetic appeal condition (through color, visual complexity, and spatial composition), the low production quality condition (through resolution), and the low aesthetic appeal plus low production quality condition, using *Adobe Photoshop*. GPT-4o was used to test whether manipulations were effective, via zero-shot questions on the aesthetic appeal and production quality on a scale of 1 to 7 (least to most appealing / lowest to highest quality).

Early results suggested that the low aesthetic appeal plus low production quality stimuli were indistinguishable from the other two conditions (low aesthetic appeal only, and low production quality only). Therefore, we decided to drop this condition and chose a partial factorial mixed design for the main experiment, with three visual conditions: original, low aesthetic appeal, and low production quality. The resulting stimuli all showed at least a 2-point difference in LLM ratings between original and low aesthetic appeal conditions, and between original and low production quality conditions (see Table B2).

**Table B2.** LLM Ratings of Aesthetic Appeal and Production Quality for Shortlisted Stimuli

| Issue Context | Aesthetic Appeal - Original | Production Quality - Original | Aesthetic Appeal - Low | Production Quality - Low |
|---|---|---|---|---|
| Climate: Photo | 6 | 7 | 4 | 3 |
| Climate: Infographic | 5 | 6 | 3 | 4 |
| Climate: Data Visualization | 5 | 6 | 4 | 4 |
| Health: Photo | 6 | 7 | 3 | 3 |
| Health: Infographic | 6 | 7 | 4 | 5 |
| Health: Data Visualization | 7 | 7 | 4 | 5 |
| RUWar: Photo | 6 | 7 | 3 | 4 |
| RUWar: Infographic | 6 | 6 | 3 | 4 |
| RUWar: Data Visualization | 6 | 6 | 4 | 4 |
| GMO: Photo | 6 | 7 | 3 | 3 |
| GMO: Infographic | 6 | 6 | 3 | 4 |
| GMO: Visualization | 5 | 6 | 3 | 4 |

*Notes*. Ratings were on a scale of 1 to 7 (lowest to highest).

**B3. Pilot Tests Employing Undergraduate Students**

After initial selection of stimuli based on LLM ratings, three more pilot tests involving undergraduate students were conducted on *Qualtrics* to ensure the manipulations were effective. The first (N=191) of these tests used an identical experimental design as the main study, randomly assigning respondents to one of the four conditions: original, low aesthetic appeal, low production quality, and text-only. Our goal for selecting satisfactory manipulated stimuli was that they should be significantly lower in aesthetic appeal and production quality, respectively, than the original visuals by one point or larger. Those stimuli not meeting this cut-off requirement were further adjusted and tested in the subsequent two rounds of testing (N = 48 and N = 20).

Due to the diverse topics and formats of stimuli tested, the manipulation differences between the original and low aesthetic appeal conditions, and between the original and low production quality conditions, were not uniform (see Table B3). This was due in part to the fact that production quality for data visualizations could not be lowered beyond a certain point, under which such visuals may become illegible. Three manipulated stimuli thus did not meet the 1-point cut-off threshold despite multiple further adjustments. Regardless, all manipulated stimuli were rated significantly lower in either aesthetic appeal or production quality than the original visuals. The average mean difference for aesthetic appeal manipulations across formats for the final selected stimuli was 1.73, and for production quality was 1.29.

**Table B3.** Pilot Test Results of Mean Differences Between High and Low Conditions for Selected Stimuli

| | Mean Difference: Original/High – Low | |
|---|---|---|
| Selected Stimuli | Aesthetic Appeal ($N_{high}$, $N_{low}$) | Production Quality ($N_{high}$, $N_{low}$) |
| Photo - Climate | $1.34^{***}$ (47, 42)[2] | $1.14^{**}$ (47, 49)[1] |
| Photo - Health | $3.90^{***}$ (42, 42)[2] | $0.92^{**}$ (47, 49)[1] |
| Photo – Russo-Ukrainian War | $2.67^{***}$ (42, 42)[2] | $0.63^{\dagger}$ (47, 49)[1] |
| Photo - GMOs | $2.12^{***}$ (47, 42)[2] | $1.04^{*}$ (47, 20)[3] |
| Infographic – Climate | $1.06^{**}$ (47, 48)[1] | $1.26^{***}$ (47, 49)[1] |
| Infographic – Health | $1.54^{***}$ (47, 48)[1] | $1.14^{*}$ (47, 20)[3] |
| Infographic – Russo-Ukrainian War | $1.49^{***}$ (47, 48)[1] | $1.00^{***}$ (47, 49)[1] |
| Infographic – GMOs | $1.48^{**}$ (47, 20)[3] | $1.96^{***}$ (47, 20)[3] |
| Data Visualizations - Climate | $1.24^{***}$ (47, 48)[1] | $2.69^{***}$ (42, 42)[2] |
| Data Visualizations - Health | $1.01^{**}$ (47, 48)[1] | $0.72^{*}$ (47, 42)[2] |
| Data Visualizations – Russo-Ukrainian War | $1.49^{***}$ (47, 48)[1] | $1.06^{**}$ (47, 42)[2] |
| Data Visualizations - GMOs | $1.45^{***}$ (47, 48)[1] | $1.86^{***}$ (47, 42)[2] |

*Notes.* The table reports pilot test results for the mean differences in ratings for aesthetic appeal and production quality between the original and low (manipulated) versions of the twelve finalized stimuli, computed as original minus low. Numbers in the parentheses represent observations for the high and low conditions, respectively, used in the Welch two-sample t-tests. The stimuli were rated by undergraduate students on a scale of 1 to 7 (lowest to highest). [1]indicates stimuli that were deemed as having achieved a significant difference in the first pilot test. [2]indicates stimuli that were selected based on the second pilot test results. [3]indicates stimuli that were selected in the third pilot test. $^{\dagger}p < .10$, $^{*}p < .05$, $^{**}p < .01$, $^{***}p < .001$.

**C. Manipulation Check Results**

In the main experiment, two manipulation check questions were included, asking participants to rate the aesthetic appeal and production quality of visuals on a scale of 1 to 7 (low to high). The mean differences between original and low aesthetic appeal and production quality conditions are shown in Table C1, and most were not significant or much smaller than those of pilot tests. Although the exact reasons for the manipulation check differences between the pilot and main experiments were unknown, we suspect that different sample characteristics may have contributed to it. We used undergraduates for pilot tests and a large heterogeneous quota sample in the main experiment. However, we think that this does not undermine our findings as the stimuli used in the main experiment were rigorously pilot tested.

**Table C1.** Manipulation Check of Aesthetic Appeal and Production Quality Manipulations in Main Experiment

| | Mean Difference: Original/High - Low | |
|---|---|---|
| Selected Stimuli | Aesthetic Appeal ($N_{high}$, $N_{low}$) | Production Quality ($N_{high}$, $N_{low}$) |
| Photo - Climate | $0.40^{*}$ (217, 225) | –0.10 (226, 225) |
| Photo - Health | $0.39^{*}$ (217, 225) | –0.08 (226, 225) |
| Photo – Russo-Ukrainian War | 0.18 (217, 225) | –0.01 (226, 225) |
| Photo - GMOs | $0.39^{*}$ (217, 225) | 0.02 (226, 225) |
| Infographic – Climate | 0.20 (217, 225) | 0.16 (226, 225) |
| Infographic – Health | 0.08 (217, 225) | 0.19 (226, 225) |
| Infographic – Russo-Ukrainian War | –0.12 (217, 225) | 0.03 (226, 225) |
| Infographic – GMOs | –0.19 (217, 225) | –0.04 (226, 225) |
| Data Visualizations - Climate | $0.30^{\dagger}$ (217, 225) | $0.28^{\dagger}$ (226, 225) |
| Data Visualizations - Health | $0.28^{\dagger}$ (217, 225) | $0.44^{**}$ (226, 225) |
| Data Visualizations – Russo-Ukrainian War | 0.18 (217, 225) | $0.31^{\dagger}$ (226, 225) |
| Data Visualizations - GMOs | $0.45^{**}$ (217, 225) | $0.40^{*}$ (226, 225) |

*Notes.* The table reports mean differences in ratings for aesthetic appeal and production quality between the original and low conditions, computed as original minus low, in the main experiment. The number of observations used in the Welch two-sample t-tests are included in the parentheses. The ratings were on a scale of 1 to 7 (lowest to highest). $^{\dagger}p < .10$, $^{*}p < .05$, $^{**}p < .01$, $^{***}p < .001$.

## D. Descriptives / Correlation

**Table D1. Correlations and Descriptive Statistics of Major Variables (N=1200)**

| | *M* | *SD* | 1 | 2 | 3 | 4 | 5 | 6 |
|---|---|---|---|---|---|---|---|---|
| 1. Perceived Credibility | 5.06 | 0.78 | | | | | | |
| 2. Processing Fluency | 5.16 | 0.78 | $.57^{***}$ | | | | | |
| 3. Age | 46.06 | 15.60 | −.03 | −.05 | | | | |
| 4. Gender: Female | 0.51 | 0.50 | $–.06^{*}$ | .05 | .04 | | | |
| 5. Digital media literacy | 3.94 | 0.73 | $.13^{***}$ | $.14^{***}$ | $–.14^{***}$ | $−.18^{***}$ | | |
| 6. Education | 4.31 | 1.41 | $.08^{**}$ | –.01 | $.06^{*}$ | −.03 | $.12^{***}$ | |
| 7. Political Conservatism | 3.72 | 1.86 | $−.10^{***}$ | $.07^{*}$ | $.14^{***}$ | −.01 | $−.09^{**}$ | $–.03^{***}$ |

*Notes.* Education was measured as an ordinal variable with response options following a ranked progression ranging from less than high school (1) to doctorate level (7). $^{*}p < .05$, $^{**}p < .01$, $^{***}p < .001$.

**E. Mixed-effects Models**

**Table E1.** Mixed-effects Models Predicting Perceived Credibility by Visual Format and Features without Covariates

| | Visual Modality (1) vs. Text (0) | | | | Visual Features: High (1) vs. Low (0) | | | |
|---|---|---|---|---|---|---|---|---|
| | Model 1: All Visuals | Model 2: Photos | Model 3: Infographics | Model 4: Data Visualizations | Model 5: All Visuals | Model 6: Photos | Model 7: Infographics | Model 8: Data Visualizations |
| | B *(SE)* | B *(SE)* | B *(SE)* | B *(SE)* | B *(SE)* | B *(SE)* | B *(SE)* | B *(SE)* |
| Intercept | 4.903***(.043) | 4.903***(.041) | 4.903***(.042) | 4.903***(.043) | 4.972***(.078) | 4.949***(.090) | 5.193***(.085) | 4.773***(.093) |
| Visual format[a] | 0.283***(.061) | 0.146*(.065) | 0.603***(.066) | 0.099 (.067) | | | | |
| Aesthetic Appeal | | | | | 0.162*(.063) | 0.014 (.074) | 0.239***(.070) | 0.232**(.076) |
| Production Quality | | | | | 0.053 (.063) | 0.086 (.074) | 0.074 (.069) | –0.002 (.076) |
| Marginal $R^2$ | 0.011 | 0.002 | 0.036 | 0.001 | 0.003 | 0.001 | 0.006 | 0.009 |
| Conditional $R^2$ | 0.263 | 0.204 | 0.257 | 0.257 | 0.309 | 0.256 | 0.271 | 0.536 |
| Observations | 7,224 | 4,832 | 4,832 | 4,832 | 10,764 | 3,588 | 3,588 | 3,588 |

*Notes.* The dummy variables included in the analyses were visual format (posts containing visuals = 1; text-only posts = 0), aesthetic appeal (high = 1, low =0), and production quality (high = 1, low = 0). The Marginal $R^2$ values represent the proportion of variance explained by the fixed effects of predictors and the Conditional $R^2$ values represent the total variance explained by both fixed and random effects (intercepts for respondents). $^{*}p < .05$, $^{**}p < .01$, $^{***}p < .001$.

**Table E2.** Mixed-effects Models Predicting Perceived Credibility by Visual Format and Features with Covariates

| | Visual Modality (1) vs. Text (0) | | | | Visual Features: High (1) vs. Low (0) | | | |
|---|---|---|---|---|---|---|---|---|
| | Model 1: All Visuals | Model 2: Photos | Model 3: Infographics | Model 4: Data Visualizations | Model 5: All Visuals | Model 6: Photos | Model 7: Infographics | Model 8: Data Visualizations |
| | B *(SE)* | B *(SE)* | B *(SE)* | B *(SE)* | B *(SE)* | B *(SE)* | B *(SE)* | B *(SE)* |
| Intercept | 4.240$^{***}$(.222) | 4.354$^{***}$(.232) | 4.131$^{***}$(.236) | 4.321$^{***}$(.241) | 4.651$^{***}$(.203) | 4.864$^{***}$(.238) | 4.610$^{***}$(.223) | 4.480$^{***}$(.242) |
| Visual format | 0.291$^{***}$(.060) | 0.153$^{*}$(.064) | 0.612$^{***}$(.065) | 0.104 (.066) | | | | |
| Aesthetic Appeal | | | | | 0.165$^{**}$(.063) | 0.015 (.074) | 0.245$^{***}$(.069) | 0.235$^{**}$(.075) |
| Production Quality | | | | | 0.049 (.062) | 0.086 (.073) | 0.068 (.069) | –0.008 (.075) |
| Age | 0.001 (.002) | 0.001 (.002) | 0.002 (.002) | –0.001 (.002) | 0.001 (.002) | 0.001 (.002) | 0.004$^{*}$(.002) | –0.001 (.002) |
| Political conservatism (ideology) | −0.031 (.017) | –0.040$^{*}$(.017) | –0.038$^{*}$(.018) | –0.020 (.018) | –0.037$^{**}$(.014) | –0.059$^{***}$(.016) | -0.041$^{**}$(.015) | –0.010 (.017) |
| Gender: Female | 0.057 (.061) | 0.097(.064) | 0.076 (.065) | 0.001 (.067) | –0.104$^{*}$(.052) | 0.019 (.061) | –0.053 (.057) | –0.277$^{***}$(.063) |
| Education: bachelor's or higher | 0.051 (.060) | 0.077 (.063) | 0.045 (.064) | 0.054 (.065) | 0.031 (.052) | 0.080 (.061) | –0.002 (.057) | 0.015 (.062) |
| Digital Literacy | 0.166$^{***}$(.043) | 0.136$^{**}$(.045) | 0.195$^{***}$(.046) | 0.167$^{***}$(.047) | 0.113$^{**}$(.036) | 0.054 (.042) | 0.149$^{***}$(.040) | 0.135$^{**}$(.043) |
| Marginal $R^2$ | 0.022 | 0.012 | 0.050 | 0.012 | 0.013 | 0.010 | 0.019 | 0.036 |
| Conditional $R^2$ | 0.265 | 0.206 | 0.260 | 0.259 | 0.310 | 0.258 | 0.272 | 0.538 |
| Observations | 7,224 | 4,832 | 4,832 | 4,832 | 10,764 | 3,588 | 3,588 | 3,588 |

*Notes.* The dummy variables included in the analyses were visual format (posts containing visuals = 1; text-only posts = 0), aesthetic appeal (high = 1, low =0), production quality (high = 1, low = 0), Gender (female = 1, male or non-binary = 0), education (bachelor's or higher = 1, associate degree or lower = 0). The Marginal $R^2$ values represent the proportion of variance explained by the fixed effects of predictors and the Conditional $R^2$ values represent the total variance explained by both fixed and random effects (intercepts for respondents). $^{*}p < .05$, $^{**}p < .01$, $^{***}p < .001$.

## F. Mediation Models

**Table F1.** Bayesian Mediation Analysis of Visual Features on Perceived Credibility via Processing Fluency

| Effect Type | Median | SD | Credible Interval | $\hat{R}$ | Bulk ESS | Tail ESS | Median | SD | Credible Interval | $\hat{R}$ | Bulk ESS | Tail ESS |
|---|---|---|---|---|---|---|---|---|---|---|---|---|
| | (a) Aesthetic Appeal: All Visuals (obs.=7,128) | | | | | | (e) Production Quality: All Visuals (obs.=7,224) | | | | | |
| Indirect | 0.039 | 0.019 | [0.008, 0.071] | 1.00 | 5065 | 8440 | 0.046 | 0.018 | [0.017, 0.075] | 1.00 | 6007 | 6007 |
| Direct | 0.128 | 0.054 | [0.038, 0.217] | 1.00 | 2771 | 4941 | 0.005 | 0.054 | [–0.083, 0.094] | 1.00 | 3917 | 3917 |
| Total | 0.167 | 0.057 | [0.074, 0.261] | 1.00 | 2807 | 5191 | 0.051 | 0.056 | [–0.042, 0.145] | 1.00 | 3926 | 3926 |
| | (b) Aesthetic Appeal: Photos (obs.=2,376) | | | | | | (f) Production Quality: Photos (obs.=2,408) | | | | | |
| Indirect | 0.026 | 0.039 | [–0.037, 0.090] | 1.00 | 5337 | 8076 | 0.117 | 0.038 | [0.055, 0.180] | 1.00 | 6192 | 9355 |
| Direct | –0.010 | 0.063 | [–0.113, 0.092] | 1.00 | 7868 | 9944 | –0.028 | 0.061 | [–0.128, 0.072] | 1.00 | 9738 | 10942 |
| Total | 0.016 | 0.073 | [–0.105, 0.137] | 1.00 | 7570 | 10109 | 0.090 | 0.072 | [–0.028, 0.207] | 1.00 | 8361 | 10618 |
| | (c) Aesthetic Appeal: Infographics (obs.=2,376) | | | | | | (g) Production Quality: Infographics (obs.=2,408) | | | | | |
| Indirect | 0.212 | 0.029 | [0.166, 0.260] | 1.00 | 7550 | 9732 | 0.037 | 0.015 | [0.013, 0.062] | 1.00 | 7714 | 10372 |
| Direct | 0.035 | 0.062 | [–0.066, 0.138] | 1.00 | 6883 | 9471 | 0.034 | 0.061 | [–0.065, 0.134] | 1.00 | 10050 | 10417 |
| Total | 0.248 | 0.066 | [0.139, 0.359] | 1.00 | 6870 | 9273 | 0.071 | 0.063 | [–0.032, 0.174] | 1.00 | 9747 | 10021 |
| | (d) Aesthetic Appeal: Data Visualizations (obs.=2,376) | | | | | | (h) Production Quality: Data Visualizations (obs.=2,408) | | | | | |
| Indirect | –0.053 | 0.025 | [–0.095, –0.011] | 1.00 | 6138 | 9011 | 0.037 | 0.030 | [–0.012, 0.086] | 1.00 | 7843 | 10242 |
| Direct | 0.290 | 0.065 | [0.184, 0.398] | 1.00 | 3785 | 6213 | –0.042 | 0.064 | [–0.147, 0.064] | 1.00 | 5814 | 8212 |
| Total | 0.238 | 0.070 | [0.122, 0.355] | 1.00 | 4044 | 6786 | –0.005 | 0.071 | [–0.120, 0.112] | 1.00 | 5986 | 7735 |

*Notes.* The table presents posterior summaries of Bayesian mediation analyses including posterior medians, standard deviations (SD), and 90% credible intervals. Estimation was based on four Markov Chain Monte Carlo (MCMC) chains with minimum 8000 iterations per chain and weakly informative priors. Effects are considered credible when the credible intervals exclude zero. $\hat{R}$ values of 1.00 indicate model convergence. Bulk and tail effective sample sizes (ESS) reflect sampling efficiency. Coefficients are unstandardized.

**Table F2.** Bayesian Mediation Analysis of Visual Formats on Perceived Credibility via Processing Fluency

| Effect Type | Mean | SD | Credible Interval | R̂ | Bulk ESS | Tail ESS |
|---|---|---|---|---|---|---|
| | (a) Overall Visuals vs. Text (obs.=14,400) | | | | | |
| Indirect | –0.028 | 0.018 | [–0.057, 0.001] | 1.00 | 11462 | 11815 |
| Direct | 0.246 | 0.043 | [0.175, 0.316] | 1.00 | 7089 | 10251 |
| Total | 0.218 | 0.046 | [0.142, 0.294] | 1.00 | 7279 | 10133 |
| | (b) Photos vs. Text (obs.=4,832) | | | | | |
| Indirect | 0.235 | 0.031 | [0.184, 0.286] | 1.00 | 7829 | 10178 |
| Direct | –0.082 | 0.053 | [–0.170, 0.006] | 1.00 | 6164 | 9300 |
| Total | 0.153 | 0.061 | [0.053, 0.255] | 1.00 | 6409 | 9078 |
| | (c) Infographics vs. Text (obs.=4,832) | | | | | |
| Indirect | 0.256 | 0.029 | [0.209, 0.304] | 1.00 | 8426 | 10850 |
| Direct | 0.356 | 0.055 | [0.266, 0.445] | 1.00 | 5545 | 8896 |
| Total | 0.612 | 0.061 | [0.512, 0.712] | 1.00 | 6024 | 9665 |
| | (d) Data Visualizations vs. Text (obs.=4,832) | | | | | |
| Indirect | –0.429 | 0.034 | [–0.485, –0.372] | 1.00 | 6802 | 9709 |
| Direct | 0.535 | 0.056 | [0.443, 0.627] | 1.00 | 4769 | 7944 |
| Total | 0.106 | 0.064 | [0.0005, 0.212] | 1.00 | 5038 | 8388 |

*Notes.* The table presents posterior summaries of Bayesian mediation analyses including posterior medians, standard deviations (SD), and 90% credible intervals. Estimation was based on four Markov Chain Monte Carlo (MCMC) chains with minimum 8000 iterations per chain and weakly informative priors. Effects are considered credible when the credible intervals exclude zero. R̂ values of 1.00 indicate model convergence. Bulk and tail effective sample sizes (ESS) reflect sampling efficiency. Coefficients are unstandardized.